# Smooth-Wall Boundary Conditions for

# Dissipation-Based Turbulence Models[*]

D. F. Hunsaker[a], W. F. Phillips[b], and R. E. Spall[c]

[a] Corresponding Author; Assistant Professor, Department of Mechanical and Aerospace Engineering, Utah State University, 4130 Old Main Hill, Logan, Utah 84322-4130, USA; Tel: 435-797-8404; E-mail address: doug.hunsaker@usu.edu

[b] Emeritus Professor, Department of Mechanical and Aerospace Engineering, Utah State University, 4130 Old Main Hill, Logan, Utah 84322-4130, USA

[c] Professor, Department of Mechanical and Aerospace Engineering, Utah State University, 4130 Old Main Hill, Logan, Utah 84322-4130, USA

## Abstract

It is shown that the smooth-wall boundary conditions specified for commonly used dissipation-based turbulence models are mathematically incorrect. It is demonstrated that when these traditional wall boundary conditions are used, the resulting formulations allow either an infinite number of solutions or no solution. Furthermore, these solutions do not enforce energy conservation and they do not properly enforce the no-slip condition at a smooth surface. This is true for all dissipation-based turbulence models, including the $k$-$\varepsilon$, $k$-$\omega$, and $k$-$\zeta$ models. Physically correct wall boundary conditions must force both $k$ and its gradient to zero at a smooth wall. Enforcing these two boundary conditions on $k$ is sufficient to determine a unique solution to the coupled system of differential transport equations. There is no need to impose any wall boundary condition on $\varepsilon$, $\omega$, or $\zeta$ at a smooth surface and it is incorrect to do so. The behavior of $\varepsilon$, $\omega$, or $\zeta$ approaching a smooth surface is that required to satisfy the differential equations and force both $k$ and its gradient to zero at the wall.

## Nomenclature

$C_1 - C_5$ = arbitrary constants of integration, Eq. (43)

$C_{\varepsilon 1}, C_{\varepsilon 2}$ = turbulence model closure coefficients, Eqs. (6) and (7)

$C_{\omega 1}, C_{\omega 2}$ = turbulence model closure coefficients, Eqs. (73) and (74)

$C_\mu$ = turbulence model closure coefficient, Eqs. (2) and (7)

$f_1, f_2$ = wall damping functions, Eq. (9) or (77)

$f_k$ = wall damping function, Eq. (76)

$f_\mu$ = wall damping function, Eq. (8) or (75)

$E$ = wall damping function, Eq. (9)

$E^+$ = wall-scaled dimensionless wall damping function, $E^+ \equiv \nu^2 E / u_\tau^6$

$\bar{\bar{\mathbf{J}}}$ = Jacobian tensor for a vector field

$k$ = turbulent kinetic energy per unit mass, Eq. (1)

$k^+$ = wall-scaled dimensionless turbulent kinetic energy, Eq. (32)

$\hat{k}$ = arbitrary dependent variable, Eqs. (42) and (88)

$l$ = channel half width

$l^+$ = wall-scaled dimensionless channel half width, $l^+ \equiv u_\tau l / \nu$

$\hat{\bar{p}}$ = pseudo mean pressure, $\hat{\bar{p}} \equiv \bar{p} + \rho g_o Z + \frac{2}{3}\rho[k + (\nu + \nu_t)\nabla \cdot \bar{\mathbf{V}}]$, where $\bar{p}$ is the mean pressure, $g_o$ is the standard acceleration of gravity at sea level, $Z = R_E H/(R_E + H)$, $H$ is geometric altitude, and $R_E$ is the radius of the Earth.

$p^+$ = wall-scaled dimensionless pseudo mean pressure gradient, Eq. (32)

$q^+$ = change of variables, Eq. (51)

$R_t$ = turbulent dissipation Reynolds number, Eq. (35) or (81)

$R_y$ = turbulent wall Reynolds number, Eq. (35)

$\bar{\bar{\mathbf{S}}}$ = strain-rate tensor for a vector field

$u^+$ = wall-scaled dimensionless $x$-velocity component, Eq. (32)

$\hat{u}$ = arbitrary dependent variable, Eqs. (42) and (88)

$u_\tau$ = friction velocity, Eq. (28)

$\bar{\mathbf{V}}$ = mean velocity vector

$\widetilde{\mathbf{V}}$      =    fluctuating velocity vector

$\widetilde{V}$      =    magnitude of the fluctuating velocity vector

$\overline{V}_x$      =    $x$ component of mean velocity vector

$\overline{V}_y$      =    $y$ component of mean velocity vector

$\widetilde{V}_x$      =    $x$ component of fluctuating velocity vector

$x$      =    axial coordinate

$y$      =    normal coordinate measured outward from a wall

$y^+$      =    wall-scaled dimensionless $y$ coordinate, Eq. (32)

$\varepsilon$      =    turbulent energy-dissipation parameter, Eq. (1)

$\widetilde{\varepsilon}$      =    turbulent energy-dissipation parameter, $\widetilde{\varepsilon} \equiv \varepsilon - \varepsilon_o$

$\varepsilon^+$      =    wall-scaled dimensionless turbulent dissipation parameter, Eq. (32)

$\hat{\varepsilon}$      =    arbitrary dependent variable, Eq. (42)

$\varepsilon_o$      =    wall damping function, Eq. (10)

$\varepsilon_o^+$      =    wall-scaled dimensionless wall damping function, Eq. (32)

$\zeta$      =    turbulent energy-dissipation parameter, $\zeta \equiv \varepsilon / \nu$

$\theta^+$      =    change of variables, Eq. (51)

$\nu$      =    kinematic molecular viscosity

$\nu_t$      =    kinematic eddy viscosity

$\nu^+$      =    ratio of the turbulent eddy viscosity to the molecular viscosity, Eq. (33)

$\rho$      =    fluid density

$\sigma_k$      =    turbulence model closure coefficient, Eqs. (5) and (7)

$\sigma_\varepsilon$      =    turbulence model closure coefficient, Eqs. (6) and (7)

$\tau_w$      =    wall shear stress, Eq. (28)

$\omega$      =    turbulent energy-dissipation frequency, $\omega \equiv \varepsilon / (C_\mu k)$

$\omega^+$      =    wall-scaled dimensionless turbulent dissipation frequency, Eq. (79)

$\hat{\omega}$      =    arbitrary dependent variable, Eq. (88)

## 1. Introduction

Many of the turbulence models that are now commonly used for computational fluid dynamics (CFD) are based on the analogy between molecular and turbulent transport that was first proposed by Boussinesq [1]. The majority of these turbulence models are usually classified as either $k$-$\varepsilon$, $k$-$\omega$, or $k$-$\zeta$ models. Conventional $k$-$\varepsilon$, $k$-$\omega$, and $k$-$\zeta$ turbulence models are often thought of as being fundamentally different. Yet, in a larger sense, these three model classifications could all be thought of as energy-dissipation models. This is because all such models are based on the hypothesis that Boussinesq's eddy viscosity is proportional to the product of the root mean square fluctuating velocity, or $k^{1/2}$, and the dissipation length scale $k^{3/2}/\varepsilon$. The parameters $k$ and $\varepsilon$ are defined in terms of the fluctuating velocity as

$$k \equiv \tfrac{1}{2}\,\overline{\widetilde{\mathbf{V}}\cdot\widetilde{\mathbf{V}}} = \tfrac{1}{2}\,\overline{\widetilde{V'^2}}, \quad \varepsilon \equiv \nu\,\overline{\overline{\overline{\mathbf{J}(\widetilde{\mathbf{V}})}}\cdot\overline{\overline{\mathbf{J}(\widetilde{\mathbf{V}})}}} \tag{1}$$

where $\widetilde{\mathbf{V}}$ is the fluctuating velocity vector, $\overline{\overline{\mathbf{J}}}(\widetilde{\mathbf{V}})$ is its Jacobian tensor, and the overscore denotes an ensemble mean.

The eddy-viscosity model that is the foundation for all commonly used $k$-$\varepsilon$, $k$-$\omega$, and $k$-$\zeta$ turbulence models is

$$\nu_t = C_\mu\,k^2/\varepsilon \tag{2}$$

where $C_\mu$ is a dimensionless closure coefficient that is nearly universally accepted as being equal to 0.09. The $k$-$\varepsilon$ turbulence models use Eq. (2) directly. The $k$-$\omega$ turbulence models use the change of variables $\omega \equiv \varepsilon/(C_\mu k)$ to transform Eq. (2) to the equivalent relation given by $\nu_t = k/\omega$. Similarly, the $k$-$\zeta$ turbulence models use the change of variables $\zeta \equiv \varepsilon/\nu$ to transform Eq. (2) to its $k$-$\zeta$ equivalent, $\nu_t = C_\mu k^2/(\nu\zeta)$. The commonly used $k$-$\varepsilon$, $k$-$\omega$, and $k$-$\zeta$ turbulence models are all based on the hypothesis that the characteristic length scale for turbulent transport is proportional to the characteristic length scale for turbulent energy dissipation.

The $k$-$\varepsilon$ turbulence model that is the foundation for most modern Boussinesq-based turbulence models is that of Jones and Launder [2]. In addition to the algebraic equation for the kinematic eddy viscosity that is given by Eq. (2), the Jones-Launder turbulence model comprises the following equations for steady incompressible flow. The continuity equation,

$$\nabla\cdot\overline{\mathbf{V}} = 0 \tag{3}$$

the Boussinesq-based Reynolds-averaged-Navier-Stokes (RANS) equations,

$$(\overline{\mathbf{V}}\cdot\nabla)\overline{\mathbf{V}} = -\nabla\hat{\bar{p}}/\rho + \nabla\cdot[2(\nu+\nu_t)\overline{\overline{\mathbf{S}}}(\overline{\mathbf{V}})] \tag{4}$$

the Boussinesq-based turbulent-energy-transport equation,

$$\overline{\mathbf{V}} \cdot \nabla k = 2\nu_t \overline{\overline{\mathbf{S}}}(\overline{\mathbf{V}}) \cdot \overline{\overline{\mathbf{S}}}(\overline{\mathbf{V}}) - \varepsilon + \nabla \cdot [(\nu + \nu_t/\sigma_k)\nabla k] \tag{5}$$

and a turbulent-dissipation-transport equation obtained by analogy with Eq. (5)

$$\overline{\mathbf{V}} \cdot \nabla \varepsilon = 2C_{\varepsilon 1} \nu_t \frac{\varepsilon}{k} \overline{\overline{\mathbf{S}}}(\overline{\mathbf{V}}) \cdot \overline{\overline{\mathbf{S}}}(\overline{\mathbf{V}}) - C_{\varepsilon 2} \frac{\varepsilon^2}{k} + \nabla \cdot [(\nu + \nu_t/\sigma_\varepsilon)\nabla \varepsilon] \tag{6}$$

The commonly used closure coefficients for this model are

$$C_\mu = 0.09, \quad C_{\varepsilon 1} = 1.44, \quad C_{\varepsilon 2} = 1.92, \quad \sigma_k = 1.0, \quad \sigma_\varepsilon = 1.3 \tag{7}$$

In this form, the Jones-Launder $k$-$\varepsilon$ turbulence model does not exhibit the proper behavior near a solid wall. Near a no-slip boundary the turbulent velocity fluctuations and turbulent transport are suppressed by the proximity of the solid surface. Accurately modeling this suppression is critical to obtaining accurate predictions for the wall shear stress and heat transfer.

In the attempt to provide realistic results near a wall, the Jones-Launder $k$-$\varepsilon$ model is often implemented with the incorporation of what are called wall damping functions. In a general form, these wall damping functions are added to Eq. (2), Eq. (6), and the definition of $\varepsilon$,

$$\nu_t = C_\mu f_\mu k^2 / \widetilde{\varepsilon} \tag{8}$$

$$\overline{\mathbf{V}} \cdot \nabla \widetilde{\varepsilon} = 2C_{\varepsilon 1} f_1 \nu_t \frac{\widetilde{\varepsilon}}{k} \overline{\overline{\mathbf{S}}}(\overline{\mathbf{V}}) \cdot \overline{\overline{\mathbf{S}}}(\overline{\mathbf{V}}) - C_{\varepsilon 2} f_2 \frac{\widetilde{\varepsilon}^2}{k} + E + \nabla \cdot [(\nu + \nu_t/\sigma_\varepsilon)\nabla \widetilde{\varepsilon}] \tag{9}$$

$$\varepsilon = \widetilde{\varepsilon} + \varepsilon_o \tag{10}$$

A variety of $k$-$\varepsilon$ turbulence models have been proposed, which differ only in the form of the wall damping functions $f_\mu$, $f_1$, $f_2$, $E$, and $\varepsilon_o$. To complete any $k$-$\varepsilon$ model of this form, the wall damping functions are specified as prescribed functions of $\nu$, $\overline{\mathbf{V}}$, $k$, $\widetilde{\varepsilon}$, and the normal coordinate $y$, measured outward from the wall. These wall damping functions are simply empirical corrections that are added to force the model to agree more closely with experimental data.

For steady, incompressible, 2-D flow in Cartesian coordinates, the $k$-$\varepsilon$ turbulence model with wall damping functions is specified by

$$\nu_t = C_\mu f_\mu k^2 / \widetilde{\varepsilon} \tag{11}$$

$$\frac{\partial \overline{V_x}}{\partial x} + \frac{\partial \overline{V_y}}{\partial y} = 0 \tag{12}$$

$$\overline{V_x}\frac{\partial \overline{V_x}}{\partial x} + \overline{V_y}\frac{\partial \overline{V_x}}{\partial y} = -\frac{1}{\rho}\frac{\partial \hat{\overline{p}}}{\partial x} + \frac{\partial}{\partial x}\left[2(\nu+\nu_t)\frac{\partial \overline{V_x}}{\partial x}\right] + \frac{\partial}{\partial y}\left[(\nu+\nu_t)\left(\frac{\partial \overline{V_y}}{\partial x}+\frac{\partial \overline{V_x}}{\partial y}\right)\right] \tag{13}$$

$$\overline{V_x}\frac{\partial \overline{V_y}}{\partial x} + \overline{V_y}\frac{\partial \overline{V_y}}{\partial y} = -\frac{1}{\rho}\frac{\partial \hat{\overline{p}}}{\partial y} + \frac{\partial}{\partial x}\left[(\nu+\nu_t)\left(\frac{\partial \overline{V_x}}{\partial y}+\frac{\partial \overline{V_y}}{\partial x}\right)\right] + \frac{\partial}{\partial y}\left[2(\nu+\nu_t)\frac{\partial \overline{V_y}}{\partial y}\right] \tag{14}$$

$$\overline{V_x}\frac{\partial k}{\partial x} + \overline{V_y}\frac{\partial k}{\partial y} = \nu_t\left[2\left(\frac{\partial \overline{V_x}}{\partial x}\right)^2 + 2\left(\frac{\partial \overline{V_y}}{\partial y}\right)^2 + \left(\frac{\partial \overline{V_y}}{\partial x}+\frac{\partial \overline{V_x}}{\partial y}\right)^2\right]$$
$$-\tilde{\varepsilon} - \varepsilon_o + \frac{\partial}{\partial x}\left[(\nu+\nu_t/\sigma_k)\frac{\partial k}{\partial x}\right] + \frac{\partial}{\partial y}\left[(\nu+\nu_t/\sigma_k)\frac{\partial k}{\partial y}\right] \tag{15}$$

$$\overline{V_x}\frac{\partial \tilde{\varepsilon}}{\partial x} + \overline{V_y}\frac{\partial \tilde{\varepsilon}}{\partial y} = C_{\varepsilon 1}f_1\nu_t\frac{\tilde{\varepsilon}}{k}\left[2\left(\frac{\partial \overline{V_x}}{\partial x}\right)^2 + 2\left(\frac{\partial \overline{V_y}}{\partial y}\right)^2 + \left(\frac{\partial \overline{V_y}}{\partial x}+\frac{\partial \overline{V_x}}{\partial y}\right)^2\right]$$
$$-C_{\varepsilon 2}f_2\frac{\tilde{\varepsilon}^2}{k} + E + \frac{\partial}{\partial x}\left[(\nu+\nu_t/\sigma_\varepsilon)\frac{\partial \tilde{\varepsilon}}{\partial x}\right] + \frac{\partial}{\partial y}\left[(\nu+\nu_t/\sigma_\varepsilon)\frac{\partial \tilde{\varepsilon}}{\partial y}\right] \tag{16}$$

If $y$ is the normal coordinate measured outward from a smooth wall, then the traditional no-slip wall boundary conditions for $\overline{V_x}$ and $\overline{V_y}$ are

$$\overline{V_x}(x,0)=0, \quad \overline{V_y}(x,0)=0 \tag{17}$$

Likewise, the obvious no-slip wall boundary condition for $k$ at a smooth wall is

$$k(x,0)=0 \tag{18}$$

The remaining wall boundary condition is assumed to be model dependent and it is the topic of the present paper.

The mean velocity and turbulent fluctuations vanish at all points on a smooth wall. Hence, near a smooth wall, changes in the mean velocity and turbulence variables with respect to $x$ are small compared to changes with respect to $y$, and the near-wall formulation reduces to

$$\frac{\partial \overline{V_y}}{\partial y} \cong 0 \tag{19}$$

$$\overline{V_y}\frac{\partial \overline{V_x}}{\partial y} \cong -\frac{1}{\rho}\frac{\partial \hat{\overline{p}}}{\partial x} + \frac{\partial}{\partial y}\left[(\nu+\nu_t)\frac{\partial \overline{V_x}}{\partial y}\right] \tag{20}$$

$$\overline{V_y}\frac{\partial \overline{V_y}}{\partial y} \cong -\frac{1}{\rho}\frac{\partial \hat{\overline{p}}}{\partial y} + \frac{\partial}{\partial y}\left[2(\nu+\nu_t)\frac{\partial \overline{V_y}}{\partial y}\right] \tag{21}$$

$$\overline{V}_y \frac{\partial k}{\partial y} \cong \nu_t \left[ 2 \left( \frac{\partial \overline{V}_y}{\partial y} \right)^2 + \left( \frac{\partial \overline{V}_x}{\partial y} \right)^2 \right] - \widetilde{\varepsilon} - \varepsilon_o + \frac{\partial}{\partial y} \left[ (\nu + \nu_t / \sigma_k) \frac{\partial k}{\partial y} \right] \tag{22}$$

$$\overline{V}_y \frac{\partial \widetilde{\varepsilon}}{\partial y} \cong C_{\varepsilon 1} f_1 \nu_t \frac{\widetilde{\varepsilon}}{k} \left[ 2 \left( \frac{\partial \overline{V}_y}{\partial y} \right)^2 + \left( \frac{\partial \overline{V}_x}{\partial y} \right)^2 \right] - C_{\varepsilon 2} f_2 \frac{\widetilde{\varepsilon}^2}{k} + E + \frac{\partial}{\partial y} \left[ (\nu + \nu_t / \sigma_\varepsilon) \frac{\partial \widetilde{\varepsilon}}{\partial y} \right] \tag{23}$$

Integrating Eq. (19) and applying Eq. (17) yields

$$\overline{V}_y \cong 0 \tag{24}$$

Using Eq. (24) in Eq. (21) yields

$$\frac{\partial \widehat{p}}{\partial y} \cong 0 \quad \text{or} \quad \widehat{p} \cong \widehat{p}(x) \tag{25}$$

Using Eqs. (24) and (25) in Eq. (20) gives

$$\frac{\partial}{\partial y} \left[ (\nu + \nu_t) \frac{\partial \overline{V}_x}{\partial y} \right] \cong \frac{1}{\rho} \frac{d\widehat{p}}{dx} \tag{26}$$

Because the right-hand side of Eq. (26) is only a function of $x$, integrating Eq. (26) from the wall to some point $y$ that is still near the wall results in

$$\left[ (\nu + \nu_t) \frac{\partial \overline{V}_x}{\partial y} \right]_{y=0}^{y} \cong \frac{1}{\rho} \frac{d\widehat{p}}{dx} y \tag{27}$$

Because the turbulent eddy viscosity is zero at a smooth wall, the left-hand side of Eq. (27) evaluated at the wall can be written in terms of either the wall shear stress or the friction velocity,

$$\nu \frac{\partial \overline{V}_x}{\partial y}(x,0) = \frac{\tau_w(x)}{\rho} \equiv u_\tau^2(x) \tag{28}$$

After using Eq. (28) in Eq. (27), the near-wall approximation for the Boussinesq-RANS equations becomes

$$(\nu + \nu_t) \frac{\partial \overline{V}_x}{\partial y} \cong u_\tau^2 + \frac{1}{\rho} \frac{d\widehat{p}}{dx} y, \qquad \overline{V}_y \cong 0 \tag{29}$$

Similarly, after applying Eq. (24) to Eqs. (22) and (23), the near-wall approximations for the $k$- and $\varepsilon$-transport equations become

$$\frac{\partial}{\partial y} \left[ (\nu + \nu_t / \sigma_k) \frac{\partial k}{\partial y} \right] \cong \widetilde{\varepsilon} + \varepsilon_o - \nu_t \left( \frac{\partial \overline{V}_x}{\partial y} \right)^2 \tag{30}$$

$$\frac{\partial}{\partial y} \left[ (\nu + \nu_t / \sigma_\varepsilon) \frac{\partial \widetilde{\varepsilon}}{\partial y} \right] \cong C_{\varepsilon 2} f_2 \frac{\widetilde{\varepsilon}^2}{k} - C_{\varepsilon 1} f_1 \nu_t \frac{\widetilde{\varepsilon}}{k} \left( \frac{\partial \overline{V}_x}{\partial y} \right)^2 - E \tag{31}$$

The near-wall formulation given by Eqs. (29)–(31) is commonly called the parallel-flow approximation. This simplification was obtained using only the approximation that changes in the mean velocity and turbulence variables with respect to $x$ are negligible compared to changes with respect to $y$. This is approximately true for any flow in the region very close to a solid surface. Furthermore, the simplifications used in obtaining Eqs. (29)–(31) hold exactly for fully developed flow in channels.

For attached flows, it is convenient to nondimensionalize the differential equations in Eqs. (29)–(31) using the traditional wall-scaled dimensionless variables as a similarity transformation

$$y^+(x,y) \equiv \frac{u_\tau(x)\,y}{\nu}, \quad u^+(y^+) \equiv \frac{\overline{V}_x(x,y)}{u_\tau(x)}, \quad k^+(y^+) \equiv \frac{k(x,y)}{u_\tau^2(x)},$$

$$\varepsilon^+(y^+) \equiv \frac{\nu\,\widetilde{\varepsilon}(x,y)}{u_\tau^4(x)}, \quad \varepsilon_o^+(y^+) \equiv \frac{\nu\,\varepsilon_o(x,y)}{u_\tau^4(x)}, \quad p^+(x) \equiv \frac{\nu}{\rho u_\tau^3(x)}\frac{d\hat{\overline{p}}}{dx}$$

(32)

Although it is not standard convention, here we shall denote the ratio of the turbulent eddy viscosity to the molecular viscosity as $\nu^+$. Thus, from Eq. (11)

$$\nu^+ \equiv \frac{\nu_t}{\nu} = C_\mu f_\mu \frac{k^2}{\nu\widetilde{\varepsilon}} = C_\mu f_\mu \frac{(k/u_\tau^2)^2}{\nu\widetilde{\varepsilon}/u_\tau^4} = C_\mu f_\mu \frac{k^{+2}}{\varepsilon^+}$$

(33)

Applying Eqs. (32) and (33) to Eqs. (25) and (29)–(31) the near-wall formulation is

$$\frac{dp^+}{dy^+} \cong 0, \quad \frac{du^+}{dy^+} \cong \frac{1+p^+y^+}{1+\nu^+}, \quad \nu^+ = C_\mu f_\mu \frac{k^{+2}}{\varepsilon^+}$$

$$\frac{d}{dy^+}\left[(1+\nu^+/\sigma_k)\frac{dk^+}{dy^+}\right] \cong \varepsilon^+ + \varepsilon_o^+ - \nu^+\left(\frac{du^+}{dy^+}\right)^2$$

$$\frac{d}{dy^+}\left[(1+\nu^+/\sigma_\varepsilon)\frac{d\varepsilon^+}{dy^+}\right] \cong C_{\varepsilon2}f_2\frac{\varepsilon^{+2}}{k^+} - C_{\varepsilon1}f_1\nu^+\frac{\varepsilon^+}{k^+}\left(\frac{du^+}{dy^+}\right)^2 - E^+$$

(34)

where $E^+ \equiv \nu^2 E/u_\tau^6$. To complete the formulation, the damping functions $f_\mu$, $f_1$, $f_2$, $E^+$, $\varepsilon_o^+$, and six boundary conditions must be specified for this coupled sixth-order system.

Although several variations for the $k$-$\varepsilon$ wall damping functions have been proposed, none have been completely successful. In the present paper we shall examine two of the commonly used models. The first $k$-$\varepsilon$ turbulence model to be considered is that developed by Lam and Bremhorst [3]. This is typical of models that use $E^+ = \varepsilon_o^+ = 0$. The second $k$-$\varepsilon$ model to be considered here is that developed by Launder and Sharma [4]. This model is typical of those that use prescribed functions for $E^+$ and $\varepsilon_o^+$, which are nonzero.

## 2. The Lam-Bremhorst $k$-$\varepsilon$ Model

For incompressible flow, the Lam-Bremhorst $k$-$\varepsilon$ turbulence model [3] uses the wall damping functions defined by

$$R_t \equiv \frac{k^2}{\nu \varepsilon} = \frac{k^2/u_\tau^4}{\nu \varepsilon/u_\tau^4} = \frac{k^{+2}}{\varepsilon^+}, \quad R_y \equiv \frac{k^{1/2}y}{\nu} = \frac{(k/u_\tau^2)^{1/2}u_\tau y}{\nu} = k^{+1/2}y^+,$$

$$f_\mu = [1 - \exp(-0.0165R_y)]^2(1 + 20.5/R_t),$$

$$f_1 = 1 + (0.05/f_\mu)^3, \quad f_2 = 1 - \exp(-R_t^2), \quad \varepsilon_o^+ = 0, \quad E^+ = 0,$$

$$C_\mu = 0.09, \quad C_{\varepsilon 1} = 1.44, \quad C_{\varepsilon 2} = 1.92, \quad \sigma_k = 1.0, \quad \sigma_\varepsilon = 1.3$$

$$(35)$$

Hence, we see that for this $k$-$\varepsilon$ model the wall damping functions can be written in terms of only the traditional wall-scaled dimensionless variables.

There has not been universal agreement regarding appropriate wall boundary conditions for the Lam-Bremhorst $k$-$\varepsilon$ model. As a specific example of how such boundary conditions affect the flow, we shall consider the case of fully developed flow in a 2-D channel of half width $l$. The parallel-flow simplifications used in obtaining Eq. (34) hold exactly for this fully developed flow. Because $y$ was defined to be the normal coordinate measured outward from the solid wall, the symmetry boundary conditions at the channel centerline require

$$\frac{du^+}{dy^+}\bigg|_{y^+=l^+} = 0, \quad \frac{dk^+}{dy^+}\bigg|_{y^+=l^+} = 0, \quad \frac{d\varepsilon^+}{dy^+}\bigg|_{y^+=l^+} = 0 \tag{36}$$

where $l^+ \equiv u_\tau l/\nu$. The first of the four differential equations in Eq. (34) can be integrated analytically and the first of the three boundary conditions in Eq. (36) can be used with the second differential equation in Eq. (34) to evaluate $p^+$, which gives

$$p^+ = -1/l^+ \tag{37}$$

Two of the three boundary conditions required at the wall are the obvious no-slip conditions in Eqs. (17) and (18). Hence, using these two boundary conditions and the two remaining symmetry boundary conditions from Eq. (36), Eqs. (34)–(37) reduce to the incomplete one-dimensional fifth-order formulation

$$\frac{du^+}{dy^+} = \frac{1 - y^+/l^+}{1 + \nu^+}, \quad \nu^+ = C_\mu f_\mu R_t$$

$$\frac{d}{dy^+}\left[(1 + \nu^+/\sigma_k)\frac{dk^+}{dy^+}\right] = \varepsilon^+ - \nu^+\left(\frac{du^+}{dy^+}\right)^2$$

$$\frac{d}{dy^+}\left[(1 + \nu^+/\sigma_\varepsilon)\frac{d\varepsilon^+}{dy^+}\right] = C_{\varepsilon 2}f_2\frac{\varepsilon^{+2}}{k^+} - C_{\varepsilon 1}f_1\,\nu^+\frac{\varepsilon^+}{k^+}\left(\frac{du^+}{dy^+}\right)^2$$

$$R_t \equiv k^{+2}/\varepsilon^+, \quad R_y \equiv k^{+1/2}y^+,$$

$$f_\mu = [1 - \exp(-0.0165R_y)]^2(1 + 20.5/R_t),$$

$$f_1 = 1 + (0.05/f_\mu)^3, \quad f_2 = 1 - \exp(-R_t^2),$$

$$C_\mu = 0.09, \quad C_{\varepsilon 1} = 1.44, \quad C_{\varepsilon 2} = 1.92, \quad \sigma_k = 1.0, \quad \sigma_\varepsilon = 1.3,$$

$$u^+(0) = 0, \quad k^+(0) = 0, \quad k^{+\prime}(l^+) = 0, \quad \varepsilon^{+\prime}(l^+) = 0$$

(38)

where the $'$ indicates a derivative with respect to $y^+$.

One additional boundary condition is needed to complete the formulation given by Eq. (38). In their original publication, Lam and Bremhorst [3] give the remaining boundary condition for Eq. (38) as

$$\varepsilon(x,0) = \nu\frac{\partial^2 k}{\partial y^2}(x,0) \quad \text{or} \quad \varepsilon^+(0) = k^{+\prime\prime}(0)$$

(39)

where the $''$ indicates a second derivative with respect to $y^+$. Although this boundary condition is commonly accepted in the literature as being the appropriate smooth-wall boundary condition for Eq. (38), it is in fact mathematically incorrect. There are an infinite number of solutions to Eq. (38), which also satisfy Eq. (39).

To see why Eq. (39) is not a viable boundary condition for Eq. (38), consider Eq. (38) in the limit as $y^+$ approaches zero. In this limit, both $R_t$ and $R_y$ go to zero and the wall damping functions and eddy viscosity near a smooth wall reduce to

$$y^+ \to 0, \quad f_\mu = \frac{20.5(0.0165)^2\varepsilon^+y^{+2}}{k^+}, \quad \nu^+ = 20.5(0.0165)^2C_\mu k^+y^{+2},$$

$$f_1 = 1 + \frac{(0.05)^3 k^{+3}}{(20.5)^3(0.0165)^6\varepsilon^{+3}y^{+6}}, \quad f_2 = \frac{k^{+4}}{\varepsilon^{+2}}$$

(40)

Using these limiting relations in the differential equations from Eq. (38) produces the near-wall system of equations, which applies in the limit as $y^+$ approaches zero,

$$y^+ \to 0, \quad \frac{du^+}{dy^+} = 1, \quad \frac{d^2k^+}{dy^{+2}} = \varepsilon^+, \quad \frac{d^2\varepsilon^+}{dy^{+2}} = \frac{-(0.05)^3C_\mu C_{\varepsilon 1}k^{+3}}{(20.5)^2(0.0165)^4\varepsilon^{+2}y^{+4}}$$

(41)

From the development of Eq. (41) it can be seen that Eq. (38) satisfies Eq. (39), independent of the fifth boundary condition. Hence, Eq. (39) does not provide the additional information required to obtain a unique solution to the indeterminate system in Eq. (38).

As a further demonstration of why Eq. (39) is not a viable boundary condition for completing the indeterminate system in Eq. (38), consider the similar system

$$\frac{d\hat{u}}{dy} = 1 - y, \quad \frac{d^2\hat{k}}{dy^2} = \hat{\varepsilon} - y^4\left(\frac{d\hat{u}}{dy}\right)^2, \quad \frac{d^2\hat{\varepsilon}}{dy^2} = y^6 - y^2\left(\frac{d\hat{u}}{dy}\right)^2,$$

$$\hat{u}(0) = 0, \quad \hat{k}(0) = 0, \quad \hat{k}'(1) = 0, \quad \hat{\varepsilon}'(1) = 0$$

(42)

This indeterminate fifth-order system of differential equations with only four boundary conditions is mathematically similar to Eq. (38), yet it is simple enough to permit obtaining a closed-form solution by direct integration. The general solution to Eq. (42) is

$$\hat{u} = C_1 + y - \frac{y^2}{2}, \qquad \hat{k} = C_2 + C_3 y + \frac{C_4 y^2}{2} + \frac{C_5 y^3}{6} - \frac{13 y^6}{360} + \frac{21 y^7}{420} - \frac{31 y^8}{1680} + \frac{y^{10}}{5040},$$

$$\hat{\varepsilon} = C_4 + C_5 y - \frac{y^4}{12} + \frac{y^5}{10} - \frac{y^6}{30} + \frac{y^8}{56}$$

(43)

After applying the four boundary conditions given in Eq. (42) to eliminate four of the five arbitrary constants, the solution in Eq. (43) yields

$$\hat{u} = y - \frac{y^2}{2}, \qquad \hat{k} = C_4 \frac{y^2 - 2y}{2} + \frac{338 y - 92 y^3 - 182 y^6 + 252 y^7 - 93 y^8 + y^{10}}{5040},$$

$$\hat{\varepsilon} = C_4 + \frac{-92 y - 70 y^4 + 84 y^5 - 28 y^6 + 15 y^8}{840}$$

(44)

As should be expected, there are an infinite number of solutions to any indeterminate fifth-order system of differential equations with only four boundary conditions, such as that specified in either Eq. (38) or Eq. (42). However, if the mathematical logic presented by Lam and Bremhorst [3] is correct, then we should be able to reduce Eq. (44) to a single unique solution by simply applying a boundary condition obtained from the second differential equation in Eq. (42) evaluated at $y = 0$. If we accept this logic, then our final boundary condition for Eq. (42) is

$$\hat{\varepsilon}(0) = \hat{k}''(0)$$

(45)

However, the reader may not be surprised to learn that applying Eq. (45) to either Eq. (43) or Eq. (44) yields only the trivial result $C_4 = C_4$. Hence, there are an infinite number of solutions to Eq. (42) that also satisfy Eq. (45).

From the discussion presented here, it should be clear that, in developing Eq. (39) as a boundary condition for Eq. (38), Lam and Bremhorst [3] used mathematical logic that is seriously flawed. Equation (39) is certainly a valid near-wall asymptote for the $k$-transport equation in Eq. (38). Thus, Eq. (39) can be used as an alternative to the $k$-transport equation for $y^+$ approaching zero, provided that it is combined with five appropriate boundary conditions. However, Eq. (39) cannot be used in lieu of one of the five boundary conditions. If valid boundary conditions could be obtained directly from the differential equations to which they apply, separate boundary conditions would not be needed to isolate a unique solution from a general solution.

If Eq. (39) is not a mathematically viable boundary condition for Eq. (38), it may be fair for a reader to ask, "How is it possible that numerical solutions to Eq. (38) have been obtained using Eq. (39) as a boundary condition?" To answer this question, we must remember that traditional CFD algorithms provide no information regarding the uniqueness of any solutions found. If multiple solutions exist, the numerical algorithm may converge on one of these solutions, but we have no assurance that the solution satisfies the physically correct wall boundary conditions, unless those boundary conditions have all been numerically enforced. It is always the user's responsibility to specify the boundary conditions appropriately, so that any solution found will be unique and physically correct.

In a review of early turbulence models, Patel, Rodi, and Scheuerer [5] point out that using Eq. (39) as a boundary condition for Eq. (38) "*is not very convenient since it involves parts of the solution of the system of coupled differential equations.*" Although Patel, Rodi, and Scheuerer [5] did not state that the boundary condition proposed by Lam and Bremhorst [3] is incorrect, they did suggest a "*more convenient boundary condition,*"

$$\frac{\partial \varepsilon}{\partial y}(x,0) = 0 \quad \text{or} \quad \varepsilon^{+'}(0) = 0 \tag{46}$$

which is also incorrect.

Durbin [6] was the first to point out that, as boundary conditions for Eq. (38), both Eq. (39) and Eq. (46) are incorrect. Durbin [6] presented the correct smooth-wall boundary conditions for Eq. (38), which are

$$\overline{V}_x(x,0) = k(x,0) = \frac{\partial k}{\partial y}(x,0) = 0 \quad \text{or} \quad u^+(0) = k^+(0) = k^{+'}(0) = 0 \tag{47}$$

With reference to the wall boundary conditions specified by Eqs. (39) and (46), Durbin [6] states, "*These conditions must violate the energy balance, and do not ensure satisfaction of conditions* (47)." With reference to the $k$ conditions specified in Eq. (47), in a later publication Durbin [7] states, "*These two conditions on k suffice to determine the solution for the coupled system of equations; there is no need to impose conditions of ε at the wall — indeed, it would be incorrect to do so.*"

To demonstrate why the smooth-wall boundary conditions given by Durbin [6] are correct, consider the implications of the no-slip boundary condition on the turbulent velocity fluctuations. At a smooth surface, both the mean and fluctuating velocity components must vanish. The definitions of $k$ and $\varepsilon$ are given by Eq. (1). Clearly, the no-slip boundary condition applied to Eq. (1) requires Eq. (18). However, because the no-slip boundary condition places no restriction on the derivative of $\tilde{V}_x$ with respect to $y$, and $\varepsilon$ depends only on the Jacobian of $\tilde{\mathbf{V}}$, physics imposes no boundary condition on $\varepsilon$. The second wall boundary condition required for the coupled fourth-order system of $k$-$\varepsilon$ transport equations is obtained by taking the gradient of $k$, which from the definition in Eq. (1) gives

$$\nabla k \equiv \nabla\left(\tfrac{1}{2}\,\overline{\tilde{V}'^2}\right) = \tfrac{1}{2}\,\overline{\nabla \tilde{V}'^2} = \overline{\tilde{V}'\nabla \tilde{V}'} \tag{48}$$

Because the no-slip boundary condition requires $\tilde{V}' = 0$ at the wall, Eq. (48) requires

$$\nabla k(x,0) = 0 \quad \text{or} \quad k^{+\prime}(0) = 0 \tag{49}$$

Hence, we see that a no-slip wall imposes two boundary conditions on $k$ and none on $\varepsilon$. This is sufficient to determine a unique solution to the coupled fourth-order system of $k$-$\varepsilon$ transport equations. There is no need to impose a wall boundary condition on $\varepsilon$, and it is incorrect to do so. The value of $\varepsilon$ at a smooth wall is that required to satisfy both Eqs. (18) and (49), as was originally pointed out by Durbin [6].

At this point it may be useful to return to our consideration of Eq. (42), which has a closed-form solution and is mathematically similar to the fifth-order system in Eq. (38). The boundary condition for Eq. (42) that is analogous to Eq. (49) is $\hat{k}'(0) = 0$. It is easily shown that applying this boundary condition to the solution of Eq. (42) that is given in Eq. (44) yields $C_4 = 169/2520$, and the complete unique solution is

$$\hat{u} = y - \frac{y^2}{2}, \qquad \hat{k} = \frac{169y^2 - 92y^3 - 182y^6 + 252y^7 - 93y^8 + y^{10}}{5040},$$
$$\hat{\varepsilon} = \frac{169 - 276y - 210y^4 + 252y^5 - 84y^6 + 45y^8}{2520} \tag{50}$$

which results in $\hat{\varepsilon}(0) = 169/2520$ and $\hat{\varepsilon}'(0) = -23/210$. On the other hand, the boundary condition for Eq. (42) that is analogous to Eq. (46) is $\hat{\varepsilon}'(0) = 0$. Applying this boundary condition to the solution of Eq. (42) that is given in Eq. (44) yields the somewhat more troubling result $-23/210 = 0$, which may inspire some concern with regard to using Eq. (46) as a boundary condition for Eq. (38).

Examination of the incomplete fifth-order system given by Eq. (42) has revealed that using $\hat{\varepsilon}(0) = \hat{k}''(0)$ as the fifth boundary condition results in an infinite number of solutions. On the other hand, Eq. (42) has no solution if $\hat{\varepsilon}'(0) = 0$ is used as the fifth boundary condition. It can be shown that the incomplete fifth-order system in Eq. (38) exhibits very similar behavior. However, solutions to Eq. (38) must be obtained numerically.

Because fully developed flow is one dimensional, a solution to Eq. (38) combined with Eq. (49) can be obtained by direct numerical integration. This permits the use of efficient high-order numerical methods such as the fourth-order Runge-Kutta algorithm. Because such solutions can be obtained quickly on very fine grids, fully developed channel flow provides an excellent benchmark for testing more computationally intensive CFD algorithms.

To facilitate direct numerical integration, the two second-order equations in Eq. (38) can be converted to four first-order equations by using the change of variables

$$q^+ \equiv -(1 + \nu^+/\sigma_k)\frac{dk^+}{dy^+}, \quad \theta^+ \equiv -(1 + \nu^+/\sigma_\varepsilon)\frac{d\varepsilon^+}{dy^+} \tag{51}$$

Combining Eq. (38) with Eq. (49), applying the change of variables given in Eq. (51), and eliminating $\nu^+$ by direct substitution provides the complete one-dimensional fifth-order formulation

$$\frac{du^+}{dy^+} = \frac{\varepsilon^+(1 - y^+/l^+)}{\varepsilon^+ + C_\mu f_\mu k^{+2}} \equiv u^{+\prime}$$

$$\frac{dk^+}{dy^+} \equiv -\frac{\sigma_k q^+ \varepsilon^+}{\sigma_k \varepsilon^+ + C_\mu f_\mu k^{+2}}$$

$$\frac{dq^+}{dy^+} = \frac{C_\mu f_\mu k^{+2} \varepsilon^+ (1 - y^+/l^+)^2}{(\varepsilon^+ + C_\mu f_\mu k^{+2})^2} - \varepsilon^+$$

$$\frac{d\varepsilon^+}{dy^+} \equiv -\frac{\sigma_\varepsilon \theta^+ \varepsilon^+}{\sigma_\varepsilon \varepsilon^+ + C_\mu f_\mu k^{+2}} \tag{52}$$

$$\frac{d\theta^+}{dy^+} = C_{\varepsilon 1} f_1 C_\mu f_\mu k^+ u^{+\prime 2} - C_{\varepsilon 2} f_2 \frac{\varepsilon^{+2}}{k^+}$$

$$R_t \equiv k^{+2}/\varepsilon^+, \quad R_y \equiv k^{+1/2} y^+,$$

$$f_{\mu 1} = [1 - \exp(-0.0165 R_y)]^2, \quad f_{\mu 2} = 1 + 20.5/R_t,$$

$$f_\mu = f_{\mu 1} f_{\mu 2}, \quad f_1 = 1 + (0.05/f_\mu)^3, \quad f_2 = 1 - \exp(-R_t^2),$$

$$u^+(0) = 0, \quad k^+(0) = 0, \quad q^+(0) = 0,$$

$$q^+(l^+) = 0, \quad \theta^+(l^+) = 0$$

It should be noted that the new variable $q^+$ is a dimensionless form of the total diffusive flux of turbulent kinetic energy $k$, which includes both molecular and turbulent diffusion. Similarly, $\theta^+$ is a dimensionless diffusive flux for $\varepsilon$. This brings to light another physical interpretation of the boundary condition given in Eq. (49), which led directly to the equivalent boundary condition in Eq. (52), i.e., $q^+(0) = 0$. With this interpretation, Eq. (49) can be viewed as a mathematical statement of the simple fact that turbulent kinetic energy cannot be diffused through a solid wall. The formulation for fully developed flow that is given by Eq. (52) requires that $q^+$ vanish at the wall and at the centerline. Thus, all of the turbulent kinetic energy that is generated within this steady flow must also be dissipated within the flow. If a boundary condition obtained from either Eq. (39) or Eq. (46) is used in place of that obtained from Eq. (49), this energy balance is not enforced. This is the origin of Durbin's statement that, "*These conditions must violate the energy balance,*" [6].

A numerical solution to the five first-order differential equations given in Eq. (52) can be obtained using fourth-order Runge-Kutta integration combined with an appropriate numerical root-finding method. Because only three of the five boundary conditions are given at $y^+ = 0$, the solution for $\varepsilon^+(0)$ and $\theta^+(0)$ must be obtained from the differential equations. The process is started with initial estimates for $\varepsilon^+(0)$ and $\theta^+(0)$. From these initial estimates, fourth-order Runge-Kutta integration can be used to obtain $q^+(l^+)$ and $\theta^+(l^+)$. The initial estimates are then refined using an appropriate numerical method until the solution is found, which corresponds to the correct centerline values $q^+(l^+) = 0$ and $\theta^+(l^+) = 0$.

A few words of caution may be in order here. Some of the terms in Eq. (52) are numerically indeterminate if $k^+ = 0$ and/or $\varepsilon^+ = 0$. Notice that a division by zero occurs in the definition of $R_t$ for $\varepsilon^+ = 0$. Thus, depending on the compiler, conditional relations may be required to enforce $f_{\mu 2} = 1$ and $f_2 = 1$ for $\varepsilon^+ \to 0$. For most compilers, Eq. (52) is numerically indeterminate for $k^+ = 0$. In this limit, both $R_t$ and $R_y$ go to zero and the eddy viscosity and wall damping functions reduce to

$$k^+ \to 0, \quad \nu^+ = 20.5(0.0165)^2 C_\mu k^+ y^{+2}, \quad f_2 \frac{\varepsilon^{+2}}{k^+} = k^{+3},$$

$$f_1 f_\mu k^+ = 20.5(0.0165)^2 \varepsilon^+ y^{+2} + \frac{(0.05)^3 \varepsilon^+(0) y^{+2}}{8(20.5)^2(0.0165)^4}$$

(53)

Hence, for the limit $k^+ \to 0$, the formulation given in Eq. (52) should be conditionally replaced with its near-wall asymptote

$$k^+ \to 0, \quad \frac{du^+}{dy^+} = \frac{1 - y^+/l^+}{1 + \nu^+} \equiv u^{+\prime}$$

$$\frac{dk^+}{dy^+} \equiv -\frac{\sigma_k \, q^+}{\sigma_k + \nu^+}$$

$$\frac{dq^+}{dy^+} = \nu^+ u^{+\prime 2} - \varepsilon^+$$

$$\frac{d\varepsilon^+}{dy^+} \equiv -\frac{\sigma_\varepsilon \, \theta^+}{\sigma_\varepsilon + \nu^+} \tag{54}$$

$$\frac{d\theta^+}{dy^+} = C_{\varepsilon 1} C_\mu F_\varepsilon u^{+\prime 2} - C_{\varepsilon 2} k^{+3}$$

$$\nu^+ = 20.5(0.0165)^2 C_\mu k^+ y^{+2},$$

$$F_\varepsilon \equiv 20.5(0.0165)^2 \varepsilon^+ y^{+2} + \frac{(0.05)^3 \varepsilon^+(0) y^{+2}}{8(20.5)^2 (0.0165)^4}$$

To demonstrate that Eq. (39) is not a valid boundary condition for completing the formulation given in Eq. (38), the results shown in Fig. 1 were obtained from Eq. (52) using randomly selected wall boundary conditions. The no-slip wall boundary conditions were used for both $u^+$ and $k^+$, but the wall boundary conditions for $q^+$, $\varepsilon^+$, and $\theta^+$, as well as the wall-scaled dimensionless half width $l^+$ were generated as listed in Fig. 1 using the "rand()" function, which generates a random number between 0.0 and 1.0. From the results presented in Fig. 1, it can be concluded that Eq. (39) is enforced directly by Eq. (38), completely independent of the boundary conditions.

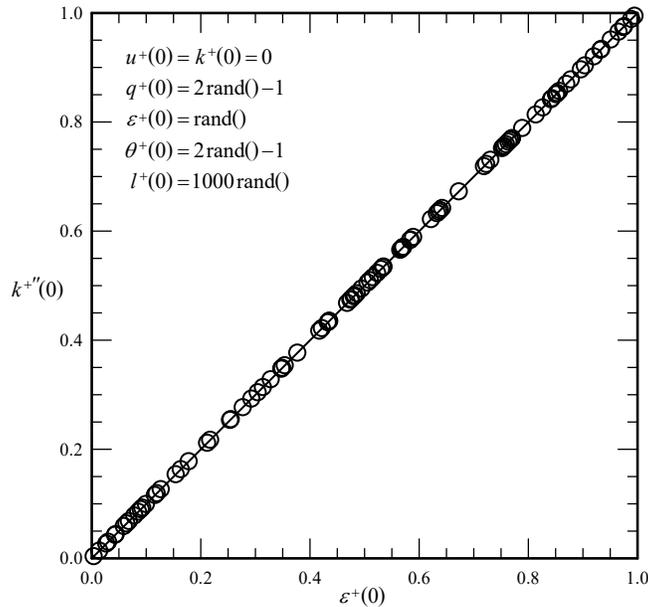

**Fig. 1  Solutions to Eq. (52) with randomly selected wall boundary conditions.**

To demonstrate that Eq. (46) is not a valid boundary condition for completing the formulation given in Eq. (38), the results shown in Fig. 2 were obtained from Eq. (52) using the no-slip wall boundary conditions for $u^+$, $k^+$, and $q^+$, with the wall boundary conditions for $\theta^+$ obtained from Eq. (46). For several values of $l^+$, the computed value for $q^+$ at the centerline is plotted as a function of the remaining wall boundary condition $\varepsilon^+(0)$. Valid solutions to Eq. (38) could only correspond to those points where these curves intersect the axis $q^+(l^+) = 0$. From the results presented in Fig. 2, it can be seen that there is only one solution to Eq. (38) that satisfies Eq. (46) and the no-slip wall boundary conditions. That is the trivial laminar solution

$$u^+ = y^+ - y^{+2}/(2l^+), \quad k^+ = q^+ = \varepsilon^+ = \theta^+ = 0 \tag{55}$$

There is no turbulent flow solution to Eq. (38) that satisfies Eq. (46) and the no-slip wall boundary conditions.

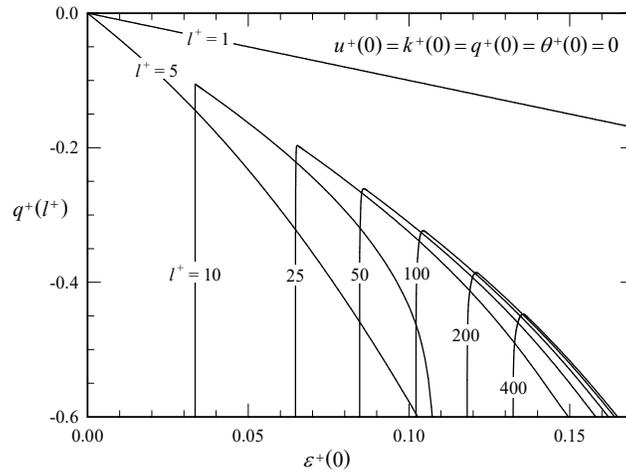

**Fig. 2   Solutions to Eq. (52) with no slip and no dissipation gradient at the wall.**

Examination of the numerical results shown in Figs. 1 and 2 reveals that Eq. (38) exhibits behavior very similar to that demonstrated analytically for the hypothetical fifth-order system given by Eq. (42). Using $\varepsilon^+(0) = k^{+''}(0)$ as the fifth boundary condition for Eq. (38) results in an infinite number of solutions. On the other hand, Eq. (38) has no valid turbulent flow solution if $\varepsilon^{+'}(0) = 0$ is used as the fifth boundary condition. This underscores the critical importance of always using the correct no-slip boundary conditions $u^+(0) = k^+(0) = k^{+'}(0) = 0$.

### 3. The Launder-Sharma $k$-$\varepsilon$ Model

For incompressible flow, the wall damping functions that are used in Eqs. (8)–(10) are defined for the Launder-Sharma $k$-$\varepsilon$ turbulence model [4] as

$$R_t \equiv k^2/(\nu\tilde{\varepsilon}), \quad f_\mu = \exp[-3.4/(1+R_t/50)^2],$$

$$f_1 = 1, \quad f_2 = 1 - 0.3\exp(-R_t^2),$$

$$\varepsilon_o = \frac{\nu}{2k}\left(\frac{\partial k}{\partial y}\right)^2, \quad E = 2\nu\nu_t\left(\frac{\partial^2 \overline{V}_x}{\partial y^2}\right)^2, \tag{56}$$

$$C_\mu = 0.09, \quad C_{\varepsilon1} = 1.44, \quad C_{\varepsilon2} = 1.92, \quad \sigma_k = 1.0, \quad \sigma_\varepsilon = 1.3$$

It may be worth noting that Launder and Sharma [4] originally defined the wall damping function $\varepsilon_o$ in a slightly different but equivalent form,

$$\varepsilon_o \equiv 2\nu\left(\frac{\partial k^{1/2}}{\partial y}\right)^2 = 2\nu\left(\tfrac{1}{2}k^{-1/2}\frac{\partial k}{\partial y}\right)^2 = \frac{\nu}{2k}\left(\frac{\partial k}{\partial y}\right)^2$$

Following what was done with the Lam-Bremhorst model, the Launder-Sharma wall damping functions can be written in terms of the wall-scaled dimensionless variables that are defined in Eqs. (32) and (33). The result applied to Eq. (34) yields the near-wall formulation for the Launder-Sharma $k$-$\varepsilon$ model

$$\frac{dp^+}{dy^+} \cong 0, \quad \frac{du^+}{dy^+} \cong \frac{1+p^+y^+}{1+\nu^+}, \quad \nu^+ = C_\mu f_\mu R_t$$

$$\frac{d}{dy^+}\left[(1+\nu^+/\sigma_k)\frac{dk^+}{dy^+}\right] \cong \varepsilon^+ + \frac{1}{2k^+}\left(\frac{dk^+}{dy^+}\right)^2 - \nu^+\left(\frac{du^+}{dy^+}\right)^2$$

$$\frac{d}{dy^+}\left[(1+\nu^+/\sigma_\varepsilon)\frac{d\varepsilon^+}{dy^+}\right] \cong C_{\varepsilon2}f_2\frac{\varepsilon^{+2}}{k^+} - C_{\varepsilon1}C_\mu f_\mu k^+\left(\frac{du^+}{dy^+}\right)^2 - 2\nu^+\left(\frac{d^2u^+}{dy^{+2}}\right)^2 \tag{57}$$

$$R_t \equiv k^{+2}/\varepsilon^+, \quad f_\mu = \exp[-3.4/(1+R_t/50)^2], \quad f_2 = 1 - 0.3\exp(-R_t^2)$$

From the first two differential equations in Eq. (57) and the specified relation for $\nu^+$, it is easily shown that the last term on the right-hand side of the $\varepsilon$-transport equation can be evaluated from

$$\frac{d^2u^+}{dy^{+2}} = \frac{p^+}{1+\nu^+} - \frac{1+p^+y^+}{(1+\nu^+)^2}\frac{d\nu^+}{dy^+} = \frac{p^+}{1+\nu^+} - \frac{1+p^+y^+}{(1+\nu^+)^2}\frac{d\nu^+}{dR_t}\frac{dR_t}{dy^+}$$

$$= \frac{p^+}{1+\nu^+} - \frac{1+p^+y^+}{(1+\nu^+)^2}\nu^+\left(1+\frac{17000\,R_t}{(50+R_t)^3}\right)\left(\frac{2}{k^+}\frac{dk^+}{dy^+} - \frac{1}{\varepsilon^+}\frac{d\varepsilon^+}{dy^+}\right) \tag{58}$$

The wall boundary conditions for this model as specified originally by Launder and Sharma [4] are

$$\overline{V}_x(x,0) = k(x,0) = \tilde{\varepsilon}(x,0) = 0 \quad \text{or} \quad u^+(0) = k^+(0) = \varepsilon^+(0) = 0 \tag{59}$$

To examine the near-wall behavior of the Launder-Sharma $k$-$\varepsilon$ model, consider the Taylor-series expansions

$$k^+(y^+) = k^+(0) + k^{+\prime}(0)y^+ + \frac{k^{+\prime\prime}(0)}{2}y^{+2} + \cdots$$
$$\varepsilon^+(y^+) = \varepsilon^+(0) + \varepsilon^{+\prime}(0)y^+ + \frac{\varepsilon^{+\prime\prime}(0)}{2}y^{+2} + \cdots \tag{60}$$

If we impose the wall boundary conditions $k^+(0) = \varepsilon^+(0) = 0$ and leave $k^{+\prime}(0)$ as an unknown constant, the near-wall expansions for the damping functions and eddy viscosity are

$$k^+(y^+) = k^{+\prime}(0)y^+ + \frac{k^{+\prime\prime}(0)}{2}y^{+2} + \cdots$$
$$\varepsilon^+(y^+) = \varepsilon^{+\prime}(0)y^+ + \frac{\varepsilon^{+\prime\prime}(0)}{2}y^{+2} + \cdots$$
$$\varepsilon_o^+ = \frac{1}{2k^+}\left(\frac{\partial k^+}{\partial y^+}\right)^2 = \frac{k^{+\prime}(0)}{2y^+} + \frac{3\,k^{+\prime\prime}(0)}{4} + \cdots \tag{61}$$
$$R_t \equiv \frac{k^{+2}}{\varepsilon^+} = \frac{k^{+\prime 2}(0)}{\varepsilon^{+\prime}(0)}y^+ + \left[\frac{k^{+\prime}(0)\,k^{+\prime\prime}(0)}{\varepsilon^{+\prime}(0)} - \frac{k^{+\prime 2}(0)\,\varepsilon^{+\prime\prime}(0)}{2\varepsilon^{+\prime 2}(0)}\right]y^{+2} + \cdots$$
$$f_\mu = \exp(-3.4) + \cdots, \quad \nu^+ = C_\mu \exp(-3.4)\frac{k^{+\prime 2}(0)}{\varepsilon^{+\prime}(0)}y^+ + \cdots, \quad f_2 = 0.7 + \cdots$$

Using these expansions in Eq. (57), the near-wall expansion for the $k$-transport equation yields

$$\frac{d^2 k^+}{dy^{+2}} = \frac{k^+(0)}{2y^+} + \frac{3\,k^{+\prime\prime}(0)}{4} - C_\mu \exp(-3.4)\frac{k^{+\prime 3}(0)}{\sigma_k\,\varepsilon^{+\prime}(0)} + \cdots \tag{62}$$

which is singular at the wall. Notice that enforcing $\varepsilon^+(0) = 0$ does not enforce $k^{+\prime}(0) = 0$. On the other hand, if we apply the wall boundary conditions $k^+(0) = k^{+\prime}(0) = 0$ and treat $\varepsilon^+(0)$ as an unknown, we obtain

$$\varepsilon_o^+ = \frac{1}{2k^+}\left(\frac{\partial k^+}{\partial y^+}\right)^2 = k^{+\prime\prime}(0) + \frac{2k^{+\prime\prime\prime}(0)}{3}y^+ + \cdots$$

and the near-wall expansion for the $k$-transport equation becomes

$$\frac{d^2 k^+}{dy^{+2}} = \varepsilon^+(0) + k^{+\prime\prime}(0) + \frac{3\varepsilon^{+\prime}(0) + 2\,k^{+\prime\prime\prime}(0)}{3}y^+ + \cdots \tag{63}$$

which requires $\varepsilon^+(0) = 0$ and $k^{+\prime\prime\prime}(0) = 3\varepsilon^{+\prime}(0)$. Hence, we see that referring to the relation $\varepsilon^+(0) = 0$ as a boundary condition for the Launder-Sharma $k$-$\varepsilon$ model is a misnomer. If the correct no-slip boundary conditions specified in Eq. (47) are applied with the Launder-Sharma $k$-$\varepsilon$ model, then $\varepsilon^+(0) = 0$ follows directly from the $k$-transport equation. In any case, the correct no-slip boundary condition $k^{+\prime}(0) = 0$ should be enforced with the Launder-Sharma $k$-$\varepsilon$ model, because the $k$-transport equation is singular at the wall for nonzero values of $k^{+\prime}(0)$. With the correct no-slip boundary conditions enforced, the near-wall expansions for the Launder-Sharma damping functions and eddy viscosity are

$$k^+(y^+) = \frac{k^{+\prime\prime}(0)}{2} y^{+2} + \frac{k^{+\prime\prime\prime}(0)}{6} y^{+3} + \cdots$$

$$\varepsilon^+(y^+) = \varepsilon^{+\prime}(0) y^+ + \frac{\varepsilon^{+\prime\prime}(0)}{2} y^{+2} + \cdots$$

$$R_t \equiv \frac{k^{+2}}{\varepsilon^+} = \frac{k^{+\prime\prime2}(0)}{4\varepsilon^{+\prime}(0)} y^{+3}$$

$$+ \left[ \frac{k^{+\prime\prime}(0) k^{+\prime\prime\prime}(0)}{6\varepsilon^{+\prime}(0)} - \frac{k^{+\prime\prime2}(0)\varepsilon^{+\prime\prime}(0)}{8\varepsilon^{+\prime2}(0)} \right] y^{+4} + \cdots \tag{64}$$

$$f_\mu = \exp(-3.4) + \cdots$$

$$\nu^+ = C_\mu \exp(-3.4) \frac{k^{+\prime\prime2}(0)}{4\varepsilon^{+\prime}(0)} y^{+3} + \cdots$$

$$f_2 = 0.7 + \cdots$$

From the development of Eq. (38), fully developed flow in a 2-D channel of half width $l$ requires $p^+ = -1/l^+$ and the Launder-Sharma model combined with the boundary conditions given in Eqs. (36) and (47) yields

$$\frac{du^+}{dy^+} = \frac{1 - y^+/l^+}{1 + \nu^+}$$

$$\frac{d}{dy^+}\left[ (1 + \nu^+/\sigma_k) \frac{dk^+}{dy^+} \right] = \varepsilon^+$$

$$+ \frac{1}{2k^+}\left( \frac{dk^+}{dy^+} \right)^2 - \nu^+ \left( \frac{du^+}{dy^+} \right)^2$$

$$\frac{d}{dy^+}\left[ (1 + \nu^+/\sigma_\varepsilon) \frac{d\varepsilon^+}{dy^+} \right] = C_{\varepsilon 2} f_2 \frac{\varepsilon^{+2}}{k^+}$$

$$- C_{\varepsilon 1} C_\mu f_\mu k^+ \left( \frac{du^+}{dy^+} \right)^2 - 2\nu^+ \left( \frac{d^2 u^+}{dy^{+2}} \right)^2$$

$$R_t \equiv k^{+2}/\varepsilon^+ \tag{65}$$

$$f_\mu = \exp[-3.4/(1 + R_t/50)^2]$$

$$\nu^+ = C_\mu f_\mu R_t$$

$$f_2 = 1 - 0.3\exp(-R_t^2)$$

$$u^+(0) = 0, \quad k^+(0) = 0, \quad k^{+\prime}(0) = 0,$$

$$k^{+\prime}(l^+) = 0, \quad \varepsilon^{+\prime}(l^+) = 0$$

Using the change of variables defined in Eq. (51) combined with Eq. (58) yields the complete one-dimensional fifth-order formulation

$$\frac{du^+}{dy^+} = \frac{1 - y^+/l^+}{1 + v^+} \equiv u^{+\prime}$$

$$\frac{dk^+}{dy^+} = -\frac{\sigma_k q^+}{\sigma_k + v^+} \equiv k^{+\prime}$$

$$\frac{dq^+}{dy^+} = v^+ u^{+\prime 2} - \varepsilon^+ - \frac{k^{+\prime 2}}{2k^+}$$

$$\frac{d\varepsilon^+}{dy^+} = -\frac{\sigma_\varepsilon \theta^+}{\sigma_\varepsilon + v^+} \equiv \varepsilon^{+\prime}$$

$$u^{+\prime\prime} \equiv \frac{-1/l^+}{1 + v^+} - \frac{v^+ u^{+\prime}}{1 + v^+}\left(1 + \frac{17000\,R_t}{(50 + R_t)^3}\right)\left(\frac{2\,k^{+\prime}}{k^+} - \frac{\varepsilon^{+\prime}}{\varepsilon^+}\right) \tag{66}$$

$$\frac{d\theta^+}{dy^+} = C_{\varepsilon 1} C_\mu f_\mu k^+ u^{+\prime 2} + 2v^+ u^{+\prime\prime 2} - C_{\varepsilon 2} f_2 \frac{\varepsilon^{+2}}{k^+}$$

$$R_t \equiv k^{+2}/\varepsilon^+, \quad f_\mu = \exp[-3.4/(1 + R_t/50)^2]$$

$$v^+ = C_\mu f_\mu R_t, \quad f_2 = 1 - 0.3\exp(-R_t^2),$$

$$u^+(0) = 0, \quad k^+(0) = 0, \quad q^+(0) = 0$$

$$q^+(l^+) = 0, \quad \theta^+(l^+) = 0$$

For the limit $y^+ \to 0$, the numerically indeterminate terms in Eq. (66) should be conditionally replaced with their near-wall asymptotes

$$y^+ \to 0, \quad R_t = 0, \quad v^+ = 0, \quad \frac{k^{+\prime 2}}{2k^+} = -q^{+\prime}(0), \quad u^{+\prime\prime} = \frac{-1/l^+}{1 + v^+}, \quad \frac{\varepsilon^{+2}}{k^+} = -\frac{2\theta^{+2}(0)}{q^{+\prime}(0)} \tag{67}$$

Note that in the limit $y^+ = 0$, the differential equation for $q^+$ reduces to the algebraic equation $\varepsilon^+(0) = 0$. Moreover, this differential equation is satisfied for any value of $q^{+\prime}(0)$. Thus, $q^{+\prime}(0)$ is an unknown constant that can be varied along with $\theta^+(0)$ to enforce the two centerline boundary conditions. If the no-slip boundary condition $q^+(0) = 0$ is not enforced, the differential formulation is mathematically indeterminate.

It should be emphasized that, because physics imposes two wall boundary conditions on $k$ and none on $\varepsilon$, the value of $\varepsilon$ throughout the flow field, including its value at the wall, must be determined exclusively from the differential transport equations while enforcing the two wall boundary conditions on $k$. Because the relation $\varepsilon^+(0) = 0$ follows directly from the differential equations as shown in Eq. (63), this is certainly a valid relation that can be used to replace a differential transport equation at the wall, provided that it is combined with a complete set of appropriate boundary conditions. However, it cannot be used in lieu of one of the required boundary conditions, as was proposed originally by Launder and Sharma [4] in their presentation of this classical turbulence model.

## 4. Numerical Results from CFD Algorithms

Because the zero-gradient boundary condition for $k$ in Eq. (47) is not explicitly enforced in many commonly implemented $k$-$\varepsilon$ turbulence models, solutions obtained from these models are not unique. To demonstrate this fact, the RANS formulations for fully developed channel flow presented in Eqs. (38) and (65) were solved numerically using a second-order finite difference algorithm with successive underrelaxation. Solutions were obtained on the domain extending from the wall to the channel centerline, and grid points were clustered near the wall using logarithmic clustering. To ensure that all results were fully converged, the successive underrelaxation was allowed to continue until the observed changes were reduced to within the double-precision machine accuracy.

To ensure that all results were grid resolved, the grids were uniformly refined until no significant changes were observed with additional grid refinement. For a given axial pressure gradient, the Launder-Sharma model required a somewhat finer grid than was required for the Lam-Bremhorst model. Results of an example grid-resolution study for the Launder-Sharma model are shown in Figs. 3−5. All results shown in these figures were obtained using the fixed axial pressure gradient, which yields a value of $y^+$ at the centerline equal to 300. For the grid refinements shown in Figs. 3−5, the four grids produced channel Reynolds numbers (based on the channel width and mean velocity) that were equal to 10,009, 10,653, 10,832, and 10,878, respectively. An additional refinement of the grid to 401 nodes, which is not shown in Fig. 3, produced a channel Reynolds number of 10,889. From these and other similar results, it was concluded that for Reynolds numbers on the order of 10,000, the 201-node grid used for Figs. 3−5 produced adequate grid resolution with both the Lam-Bremhorst and Launder-Sharma turbulence models.

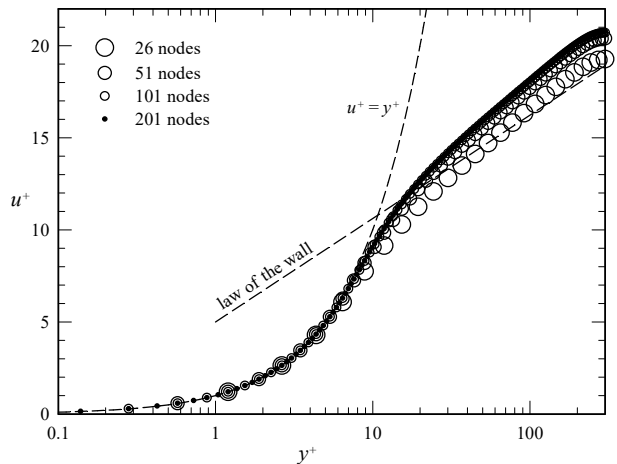

**Fig. 3  Grid resolution for the mean velocity predicted from the Launder-Sharma $k$-$\varepsilon$ model.**

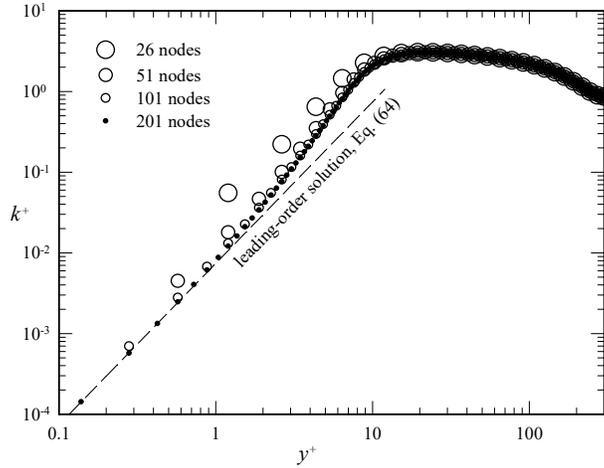

**Fig. 4   Grid resolution for the turbulent energy predicted from the Launder-Sharma $k$-$\varepsilon$ model.**

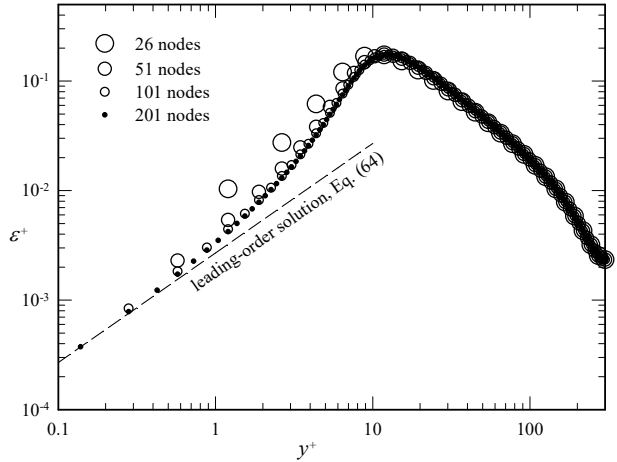

**Fig. 5   Grid resolution for the turbulent dissipation predicted from the Launder-Sharma $k$-$\varepsilon$ model.**

To demonstrate that solutions obtained from commonly implemented $k$-$\varepsilon$ turbulence models are not unique, the second-order successive underrelaxation algorithm was implemented using a slight variant of the wall boundary conditions specified in Eq. (47), which allows the user to specify an arbitrary value for the gradient of $k$ at the wall. Figures 6 and 7 show converged and grid-resolved results obtained from this algorithm using three different gradient boundary conditions for $k$ at the wall: $k^{+\prime}(0) = 0.0$, $k^{+\prime}(0) = 0.1$, and $k^{+\prime}(0) = 1.0$.

To demonstrate that traditional implementations of these turbulence models do not necessarily converge to the solution that yields $k^{+\prime}(0) = 0$, Figs. 6 and 7 also include results obtained from the same algorithm, turbulence models, and grid, but with a traditional implementation of the wall boundary conditions, which uses only $u^+(0) = 0$ and $k^+(0) = 0$ together with an asymptotic relation obtained from the differential equations, i.e., $\varepsilon^+(0) = k^{+''}(0)$ for the Lam-Bremhorst model and $\varepsilon^+(0) = 0$ for the Launder-Sharma model. For additional comparison, Figs. 6 and 7 also show results from the general-purpose finite-volume CFD solver FLUENT [8], which were obtained using the same turbulence models and grid with only the traditional wall boundary conditions implemented.

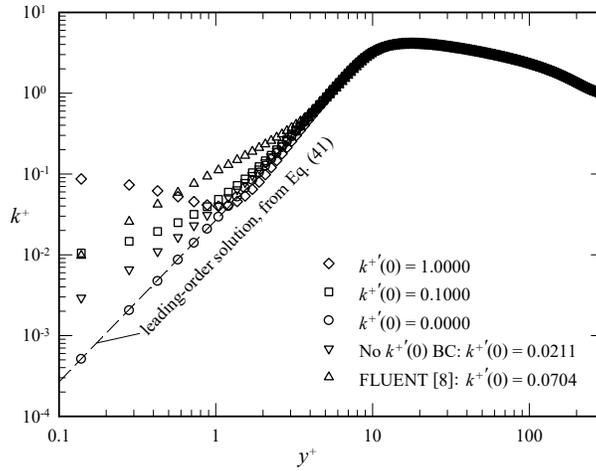

**Fig. 6   Effects of wall boundary conditions on turbulent energy predicted from the Lam-Bremhorst model.**

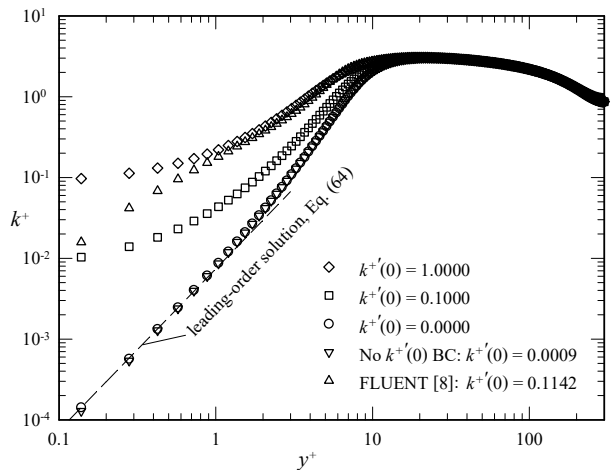

**Fig. 7   Effects of wall boundary conditions on turbulent energy predicted from the Launder-Sharma model.**

The results shown in Figs. 6 and 7 clearly demonstrate that when the natural boundary condition $k^{+'}(0) = 0$ is not enforced, solutions obtained from commonly implemented $k$-$\varepsilon$ turbulence models are not unique. When the boundary condition $k^{+'}(0) = 0$ is omitted, solutions obtained from the resulting indeterminate formulation are implementation dependent. Notice from Fig. 6 that for the Lam-Bremhorst model with traditional implementation of the wall boundary conditions, the finite-difference algorithm converged to a different solution from that obtained using the finite-volume algorithm with the same indeterminate boundary conditions. Neither of these solutions agrees with that obtained from the finite-difference algorithm with the boundary condition $k^{+'}(0) = 0$ enforced. Similarly, we see from Fig. 7 that these finite-difference and finite-volume implementations of the Launder-Sharma model converge to different solutions with traditional implementations of the wall boundary conditions. However, for the particular implementation used to obtain the results shown in Fig. 7, the indeterminate finite-difference algorithm converged to a solution that is very close to that obtained when the complete set of smooth-wall boundary conditions was enforced. This should not be viewed as an endorsement for implementing the Launder-Sharma turbulence model with mathematically incomplete boundary conditions.

From the near-wall expansions of the Launder-Sharma model given in Eqs. (62) and (63), it was shown that enforcing $k^{+'}(0) = 0$ requires $\varepsilon^{+}(0) = 0$, whereas enforcing $\varepsilon^{+}(0) = 0$ does not require $k^{+'}(0) = 0$. This can also be demonstrated numerically by examining the near-wall behavior of $\varepsilon$ obtained from numerical solutions using different gradient boundary conditions for $k$ at the wall. Such results are shown in Fig. 8, which were obtained from converged and grid-resolved solutions for the same channel flow that was used to obtain the results shown in Fig. 7. Notice that, although $k^{+'}(0)$ does affect the near-wall behavior of $\varepsilon^{+}$, all of these solutions satisfy $\varepsilon^{+}(0) = 0$. Only the solution corresponding to $k^{+'}(0) = 0$ also satisfies the physically correct no-slip condition at the smooth wall.

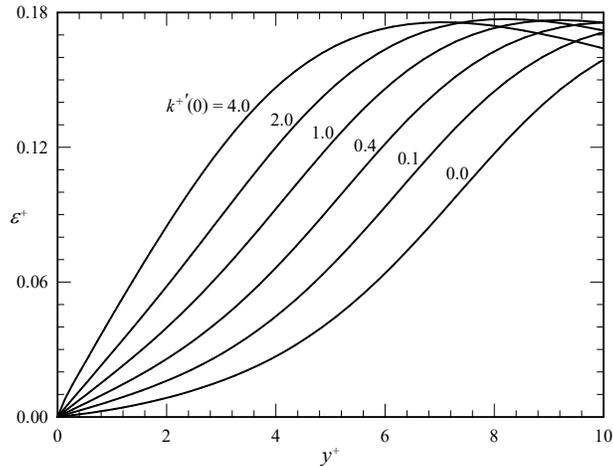

**Fig. 8   Effects of wall boundary conditions on near-wall dissipation for the Launder-Sharma model.**

Because the wall damping functions for the Lam-Bremhorst model decay rapidly with increasing $y^+$, the wall boundary condition $k^{+'}(0) = 0$ has little impact on the velocity profiles predicted from this turbulence model. On the other hand, the wall damping functions for the Launder-Sharma model decay slower and have a more significant effect on the predicted mean velocity farther from the wall. This can be seen in Fig. 9, which displays the velocity profiles for the same solutions that were used to obtain the turbulent energy profiles displayed in Fig. 7. It may be worth reiterating that all of the solutions shown in Fig. 9 satisfy the traditional wall boundary conditions $u^+(0) = 0$ and $k^+(0) = 0$ together with the asymptotic relation obtained from the differential equations, $\varepsilon^+(0) = 0$.

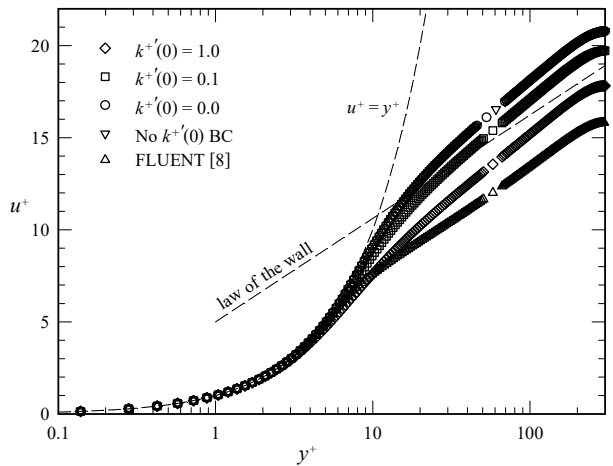

**Fig. 9** Effects of wall boundary conditions on the mean velocity predicted from the Launder-Sharma model.

Anyone who has taken time to compare results obtained from different CFD algorithms and different $k$-$\varepsilon$ turbulence models will likely have noticed that there is often a greater difference between the results obtained from two different implementations of the same turbulence model than there is between the results obtained from the same implementation of two different turbulence models [9]. The results shown in Fig. 9 may shed some light on the reason for this phenomenon. We should not be too surprised to learn that results obtained from commonly used $k$-$\varepsilon$ turbulence models are implementation dependent, if we recognize that these models are short one boundary condition, and thus are mathematically indeterminate.

Because the CFD community has not traditionally implemented two wall boundary conditions on $k$ and none on $\varepsilon$, implementation of the correct smooth-wall boundary conditions first proposed by Durbin [6] has been less than enthusiastically embraced. The actual implementation of these boundary conditions is dependent on the numerical method being used to solve the system of differential equations. However, this implementation should not be difficult using well-known methods in either finite-difference or finite-volume algorithms. For example, results presented in this section were obtained from a finite-difference algorithm. To implement the no-slip wall boundary conditions, the $k$-transport equation at any wall node was replaced with the boundary condition $k(0) = 0$. At the first node off the wall, the $k$-transport equation was replaced with a second-order finite-difference approximation for the boundary condition $k^{+\prime}(0) = 0$. Because there is no wall boundary condition on $\varepsilon$, the $\varepsilon$-transport equation at any wall node was replaced with the asymptotic relation obtained from the differential equations, i.e., $\varepsilon^{+}(0) = k^{+\prime\prime}(0)$ for the Lam-Bremhorst model and $\varepsilon^{+}(0) = 0$ for the Launder-Sharma model. The implementation of the $\varepsilon$-transport equation is identical to that of the traditional formulation. The error in the traditional formulation is not that the transport equation for $\varepsilon$ is incorrectly implemented. Rather, the error is in assuming that the near-wall asymptote obtained from the differential equations can be used to replace the final boundary condition required at the wall.

As scientists and engineers, we do not have the luxury of choosing boundary conditions for ease of numerical implementation. Boundary conditions are dictated by physics. It is our obligation to understand and implement them correctly if we hope to achieve mathematical formulations that correctly model physics. The fundamental mathematical error of deriving a so called *boundary condition* directly from the differential equations is not unique to the classical $k$-$\varepsilon$ turbulence models that have been considered here. It is also an important concern for many other turbulence models developed more recently [10–13].

## 5. Application to $k$-$\omega$ Models

The $k$-$\omega$ turbulence models that are commonly used for CFD are built on exactly the same dissipation-based eddy-viscosity model that is given in Eq. (2), where $k$ and $\varepsilon$ are defined by Eq. (1). These commonly used $k$-$\omega$ turbulence models are based on applying a simple change of variables to Eq. (2), i.e.,

$$\omega \equiv \varepsilon / (C_\mu k) \tag{68}$$

This change of variables applied to Eq. (2) yields an algebraic equation for the kinematic eddy viscosity in terms of only the turbulent kinetic energy per unit mass, $k$, and the turbulent energy-dissipation frequency, $\omega$,

$$\nu_t = k / \omega \tag{69}$$

In addition to this algebraic equation for the kinematic eddy viscosity, the $k$-$\omega$ turbulence model originally proposed by Kolmogorov [14] has been refined to comprise the following equations for steady incompressible flow. The continuity equation combined with the Boussinesq-RANS equations,

$$\nabla \cdot \overline{\mathbf{V}} = 0 \tag{70}$$

$$(\overline{\mathbf{V}} \cdot \nabla)\overline{\mathbf{V}} = -\nabla\hat{\overline{p}}/\rho + \nabla \cdot [2(\nu + \nu_t)\overline{\overline{\mathbf{S}}}(\overline{\mathbf{V}})] \tag{71}$$

the Boussinesq-based turbulent-energy-transport equation obtained by applying the change of variables defined in Eq. (68) to Eq. (5),

$$\overline{\mathbf{V}} \cdot \nabla k = 2\nu_t \overline{\overline{\mathbf{S}}}(\overline{\mathbf{V}}) \cdot \overline{\overline{\mathbf{S}}}(\overline{\mathbf{V}}) - C_\mu k\omega + \nabla \cdot [(\nu + \nu_t/\sigma_k)\nabla k] \tag{72}$$

and a dissipation-frequency-transport equation obtained by analogy with Eq. (72),

$$\overline{\mathbf{V}} \cdot \nabla \omega = 2C_{\omega 1}\nu_t \frac{\omega}{k}\overline{\overline{\mathbf{S}}}(\overline{\mathbf{V}}) \cdot \overline{\overline{\mathbf{S}}}(\overline{\mathbf{V}}) - C_{\omega 2}\omega^2 + \nabla \cdot [(\nu + \nu_t/\sigma_\omega)\nabla\omega] \tag{73}$$

The closure coefficients differ slightly from one version of the model to another and have changed as the model has evolved over the past six decades. In the original $k$-$\omega$ model, Kolmogorov [14] assumed $C_{\omega 1} = 0$ and he did not include the molecular diffusion term. The closure coefficients often used for the $k$-$\omega$ model [8,15] are

$$C_\mu = 0.09, \quad C_{\omega 1} = 0.52, \quad C_{\omega 2} = 0.072, \quad \sigma_k = 2.0, \quad \sigma_\omega = 2.0 \tag{74}$$

It should be noted that the turbulence variable $\omega$, which is defined by Eq. (68) and referred to here as the turbulent energy-dissipation frequency, is often referred to as the specific dissipation rate.

As is the case for the $k$-$\varepsilon$ model, the standard $k$-$\omega$ model does not exhibit the proper behavior near a solid wall. By direct analogy with what has been done with the $k$-$\varepsilon$ model, the $k$-$\omega$ model could also be implemented with the incorporation of wall damping functions. Although this terminology is not commonly used with the $k$-$\omega$ model, to emphasize similarities between the low-Reynolds-number corrections used for the $k$-$\omega$ model and those used for the $k$-$\varepsilon$ model, here we will use exactly the same notation and terminology for both models. Adding wall damping functions to Eqs. (69), (72), and (73) yields

$$\nu_t = f_\mu k/\omega \tag{75}$$

$$\overline{\mathbf{V}} \cdot \nabla k = 2\nu_t \overline{\overline{\mathbf{S}}}(\overline{\mathbf{V}}) \cdot \overline{\overline{\mathbf{S}}}(\overline{\mathbf{V}}) - C_\mu f_k k\omega + \nabla \cdot [(\nu + \nu_t/\sigma_k)\nabla k] \tag{76}$$

$$\overline{\mathbf{V}} \cdot \nabla \omega = 2C_{\omega 1} f_1 \nu_t \frac{\omega}{k}\overline{\overline{\mathbf{S}}}(\overline{\mathbf{V}}) \cdot \overline{\overline{\mathbf{S}}}(\overline{\mathbf{V}}) - C_{\omega 2} f_2 \omega^2 + \nabla \cdot [(\nu + \nu_t/\sigma_\omega)\nabla\omega] \tag{77}$$

To complete any $k$-$\omega$ turbulence model in this form, the wall damping functions $f_\mu$, $f_k$, $f_1$, and $f_2$, could be specified as prescribed functions of $\nu$, $\overline{\mathbf{V}}$, $k$, and $\omega$. As is the case for the $k$-$\varepsilon$ model, these wall damping functions are simply empirical corrections, which are employed to force the model to agree more closely with near-wall experimental data.

Following the development of Eq. (34) for steady, incompressible, 2-D flow in Cartesian coordinates, the near-wall approximation for the $k$-$\omega$ turbulence model with wall damping functions can be written as

$$\frac{\partial \hat{\overline{p}}}{\partial y} \cong 0, \quad \overline{V}_y \cong 0, \quad (\nu + \nu_t)\frac{\partial \overline{V}_x}{\partial y} \cong u_\tau^2 + \frac{1}{\rho}\frac{d\hat{\overline{p}}}{dx}y, \quad \nu_t = f_\mu \frac{k}{\omega}$$

$$\frac{\partial}{\partial y}\left[(\nu + \nu_t/\sigma_k)\frac{\partial k}{\partial y}\right] \cong C_\mu f_k k\omega - \nu_t\left(\frac{\partial \overline{V}_x}{\partial y}\right)^2 \tag{78}$$

$$\frac{\partial}{\partial y}\left[(\nu + \nu_t/\sigma_\omega)\frac{\partial \omega}{\partial y}\right] \cong C_{\omega 2} f_2 \omega^2 - C_{\omega 1} f_1 \nu_t \frac{\omega}{k}\left(\frac{\partial \overline{V}_x}{\partial y}\right)^2$$

Continuing to follow what was done with the $k$-$\varepsilon$ model, the differential equations in this formulation can be nondimensionalized using the wall-scaled dimensionless variables defined in Eqs. (32) and (33) together with the traditional definition for $\omega^+$

$$\omega^+(y^+) \equiv \frac{\nu \omega(x, y)}{u_\tau^2(x)} \tag{79}$$

The result yields the dimensionless near-wall $k$-$\omega$ formulation for steady 2-D incompressible flow

$$\frac{dp^+}{dy^+} \cong 0, \quad \frac{du^+}{dy^+} \cong \frac{1 + p^+ y^+}{1 + \nu^+}, \quad \nu^+ = f_\mu \frac{k^+}{\omega^+}$$

$$\frac{d}{dy^+}\left[(1 + \nu^+/\sigma_k)\frac{dk^+}{dy^+}\right] \cong C_\mu f_k k^+ \omega^+ - \nu^+\left(\frac{du^+}{dy^+}\right)^2 \tag{80}$$

$$\frac{d}{dy^+}\left[(1 + \nu^+/\sigma_\omega)\frac{d\omega^+}{dy^+}\right] \cong C_{\omega 2} f_2 \omega^{+2} - C_{\omega 1} f_1 \nu^+ \frac{\omega^+}{k^+}\left(\frac{du^+}{dy^+}\right)^2$$

As an example of a $k$-$\omega$ turbulence model that includes wall damping functions, consider what is commonly called the Wilcox 1998 $k$-$\omega$ model [15], which is implemented in FLUENT [8]. Although Wilcox [15] uses a different notation, his formulation is easily rearranged to the format of Eq. (78). For 2-D incompressible flow the resulting wall damping functions are

$$R_t \equiv \frac{k}{\nu\,\omega}, \quad \chi_k \equiv \frac{\nabla k \cdot \nabla \omega}{\omega^3}, \quad f_\mu = \frac{0.024 + R_t/6}{1 + R_t/6},$$

$$f_k = \frac{4/15 + (R_t/8)^4}{1 + (R_t/8)^4} \begin{Bmatrix} 1, & \chi_k \leq 0 \\ \dfrac{1 + 680\chi_k^2}{1 + 400\chi_k^2}, & \chi_k > 0 \end{Bmatrix}, \quad f_1 = \frac{1/9 + R_t/2.95}{f_\mu(1 + R_t/2.95)}, \quad f_2 = 1, \tag{81}$$

$$C_\mu = 0.09, \quad C_{\omega 1} = 0.52, \quad C_{\omega 2} = 0.072, \quad \sigma_k = 2.0, \quad \sigma_\omega = 2.0$$

Notice that the turbulent dissipation Reynolds number $R_t$, as it is defined in the Wilcox 1998 $k$-$\omega$ model, differs from that defined for the $k$-$\varepsilon$ models, i.e., $(R_t)_{k-\omega} \equiv C_\mu (R_t)_{k-\varepsilon}$.

Following the development of Eq. (38), these wall damping functions can be written in terms of the wall-scaled dimensionless variables that are defined in Eqs. (32) and (79). Hence, for fully developed flow in a 2-D channel of half width $l$, the Wilcox 1998 $k$-$\omega$ model combined with the boundary conditions given in Eqs. (17), (18), and (36) yields

$$\frac{du^+}{dy^+} = \frac{1 - y^+/l^+}{1 + \nu^+}, \quad \frac{d}{dy^+}\left[(1 + \nu^+/\sigma_k)\frac{dk^+}{dy^+}\right] = C_\mu f_k k^+ \omega^+ - \nu^+\left(\frac{du^+}{dy^+}\right)^2$$

$$\frac{d}{dy^+}\left[(1 + \nu^+/\sigma_\omega)\frac{d\omega^+}{dy^+}\right] = C_{\omega 2}\,\omega^{+2} - C_{\omega 1} f_1 f_\mu\left(\frac{du^+}{dy^+}\right)^2$$

$$R_t \equiv \frac{k^+}{\omega^+}, \quad \chi_k \equiv \frac{1}{\omega^{+3}}\frac{dk^+}{dy^+}\frac{d\omega^+}{dy^+}, \quad f_\mu = \frac{0.024 + R_t/6}{1 + R_t/6}, \quad \nu^+ = f_\mu R_t, \tag{82}$$

$$f_k = \frac{4/15 + (R_t/8)^4}{1 + (R_t/8)^4}\begin{Bmatrix} 1, & \chi_k \leq 0 \\ \dfrac{1 + 680\chi_k^2}{1 + 400\chi_k^2}, & \chi_k > 0 \end{Bmatrix}, \quad f_1 = \frac{1/9 + R_t/2.95}{f_\mu(1 + R_t/2.95)}$$

$$u^+(0) = 0, \quad k^+(0) = 0, \quad k^{+\prime}(l^+) = 0, \quad \omega^{+\prime}(l^+) = 0$$

As in the case of Eq. (38), one additional boundary condition is needed to complete the fifth-order formulation expressed in Eq. (82). In the presentation of his 1998 $k$-$\omega$ model Wilcox [15] states, "*The final condition follows from examination of the differential equations for k and $\omega$ approaching the surface.*" For a smooth wall in the limit $y^+ \to 0$, the boundary condition $k^+(0) = 0$ requires $R_t(0) = 0$ and $\nu^+(0) = 0$. Thus, the differential equation for $u^+$ and the $\omega$-transport equation given in Eq. (82) reduce to

$$y^+ \to 0, \quad \frac{du^+}{dy^+} = 1, \quad \frac{d^2\omega^+}{dy^{+2}} = C_{\omega 2}\,\omega^{+2} - \frac{C_{\omega 1}}{9} \tag{83}$$

Let the leading-order term in the solution for $\omega^+$ be written as

$$\omega^+(y^+) = Ay^{+a} + \cdots \tag{84}$$

where $A$ and $a$ are as yet unknown constants. Using Eq. (84) in the near-wall approximation for the $\omega$-transport equation given by Eq. (83) yields

$$a(a-1)Ay^{+\,a-2} \cong C_{\omega 2}A^2 y^{+2a} - C_{\omega 1}/9 \qquad (85)$$

Equating the exponents and coefficients of $y^+$ in the leading-order terms, this near-wall relation requires

$$a = -2, \quad A = \frac{a(a-1)}{C_{\omega 2}} = \frac{6}{C_{\omega 2}} \qquad (86)$$

Hence, after using Eq. (86) in Eq. (84), the leading-order solution for $\omega^+$ yields

$$\omega^+(y^+) \underset{y^+ \to 0}{=} \frac{6}{C_{\omega 2}\,y^{+2}} \qquad (87)$$

To minimize numerical truncation error associated with the singularity, Wilcox [16] suggests that Eq. (87) should be used in place of the $\omega$-transport equation "*for the first 7 to 10 grid points above the surface.*" Wilcox also points out that the grid must be fine enough so that "*these grid points must lie below $y^+ = 2.5$ …*" In practice, Eq. (87) is often used as the final boundary condition by applying this relation only at the first grid point off the wall [8].

Because the leading-order solution given by Eq. (87) follows exclusively from the $\omega$-transport equation with application of only the single boundary condition $k^+(0) = 0$, all solutions to Eq. (82) will exhibit this asymptotic behavior, completely independent of the fifth boundary condition that is required to obtain a unique solution to this system of equations. Equation (87) is certainly a valid asymptote for the $\omega$-transport equation in Eq. (82) near a smooth wall. Thus, Eq. (87) can be used as an alternative to the $\omega$-transport equation for $y^+$ approaching zero, provided that it is combined with five appropriate boundary conditions. However, Eq. (87) cannot be used as a substitute for one of the five required boundary conditions.

To show that Eq. (87) is not a valid boundary condition for completing the indeterminate system in Eq. (82), consider the similar system

$$\frac{d\hat{u}}{dy} = 1 - y, \quad \frac{d^2\hat{k}}{dy^2} = y^2\hat{\omega} - y^4\left(\frac{d\hat{u}}{dy}\right)^2, \quad \frac{d^2\hat{\omega}}{dy^2} = \frac{1}{y^4} - \left(\frac{d\hat{u}}{dy}\right)^2,$$

$$\hat{u}(0) = 0, \quad \hat{k}(0) = 0, \quad \hat{k}'(1) = 0, \quad \hat{\omega}'(1) = 0 \qquad (88)$$

This indeterminate fifth-order system of differential equations with only four boundary conditions is mathematically similar to Eq. (82), yet it is simple enough to yield a closed-form solution. The general solution to Eq. (88) is

$$\hat{u} = C_1 + y - \frac{y^2}{2}, \quad \hat{k} = C_2 + C_3 y + \frac{C_4 y^4}{12} + \frac{C_5 y^5}{20} + \frac{840 y^2 - 504 y^6 + 560 y^7 - 195 y^8}{10080},$$

$$\hat{\omega} = \frac{1}{6 y^2} + C_4 + C_5 y + \frac{-6 y^2 + 4 y^3 - y^4}{12} \tag{89}$$

After applying the four boundary conditions given in Eq. (88) to eliminate four of the five arbitrary constants, the solution in Eq. (89) yields

$$\hat{u} = y - \frac{y^2}{2}, \quad \hat{k} = C_4 \frac{y^4 - 4y}{12} + \frac{-2696 y + 840 y^2 + 336 y^5 - 504 y^6 + 560 y^7 - 195 y^8}{10080},$$

$$\hat{\omega} = \frac{1}{6 y^2} + C_4 + \frac{8 y - 6 y^2 + 4 y^3 - y^4}{12} \tag{90}$$

Again as should be expected, there are an infinite number of solutions to any indeterminate fifth-order system of differential equations with only four boundary conditions, such as that specified in either Eq. (82) or Eq. (88). The remaining constant of integration $C_4$ can be evaluated only by applying a mathematically appropriate boundary condition. No amount of analysis applied to the differential equations, no matter how sophisticated, will ever yield a result from which the remaining arbitrary constant in Eq. (90) can be determined.

Notice from Eq. (89) that, analogous to the result obtained from Eq. (82), the general solution for $\hat{\omega}$ approaches $y = 0$ in proportion to $y^{-2}$. From examination of either Eq. (89) or Eq. (90), it should be clear that none of the integration constants could ever be obtained by using the asymptotic behavior of $\hat{\omega}$ for $y \to 0$ as the fifth boundary condition for Eq. (88). In fact, because the behavior of $\hat{\omega}$ for $y \to 0$ depends only on the differential equations in Eq. (88), no boundary condition for $\hat{\omega}$ can be applied to Eq. (88) at $y = 0$. Likewise, because the near-wall behavior of $\omega^+$ depends only on the differential equations in Eq. (82), no wall boundary condition for $\omega^+$ can be applied to Eq. (82). The remaining boundary condition for Eq. (88) at $y = 0$ must be applied to $\hat{k}$. Similarly, the remaining wall boundary condition for Eq. (82) must be applied to $k^+$.

As presented in Eq. (47), at a smooth wall the correct no-slip boundary condition for completing the fifth-order formulation presented in Eq. (82) is $k^{+\prime}(0) = 0$. The analogous boundary condition for Eq. (88) is $\hat{k}'(0) = 0$. It is easily shown that applying this wall boundary condition to the solution of Eq. (88) that is given in Eq. (90) yields $C_4 = -337/420$, and the complete unique solution is

$$\hat{u} = y - \frac{y^2}{2}, \quad \hat{k} = \frac{840y^2 - 674y^4 + 336y^5 - 504y^6 + 560y^7 - 195y^8}{10080},$$

$$\hat{\omega} = \frac{1}{6y^2} + \frac{-337 + 280y - 210y^2 + 140y^3 - 35y^4}{420}$$

(91)

Hence, we see that imposing two wall boundary conditions on $\hat{k}$ and none on $\hat{\omega}$ is sufficient to determine a unique solution to the coupled fifth-order system of differential equations given in Eq. (88). There is no need to impose a wall boundary condition on $\hat{\omega}$, and it is incorrect to do so.

It can be shown numerically that the Wilcox 1998 $k$-$\omega$ formulation given in Eq. (82) exhibits behavior similar to that shown analytically for Eq. (88). For example, Figs. 10 and 11 show $k^+$ and $\omega^+$ for five solutions, which all satisfy both Eqs. (82) and (87). These converged and grid-resolved solutions were obtained using the same second-order successive underrelaxation algorithms that were used to obtain the $k$-$\varepsilon$ solutions shown in Figs. 6 and 7. These results demonstrate that it is mathematically incorrect to use Eq. (87) as the sole substitute for the remaining boundary condition, which is required to obtain a unique solution to Eq. (82). Neither Eq. (87) nor any other relation obtained solely from the differential equations can be used to obtain a unique solution from the indeterminate $k$-$\omega$ formulation given in Eq. (82). In addition to using Eq. (87) for numerical implementation, all three of the no-slip boundary conditions that are given in Eq. (47) should be explicitly enforced at a smooth wall.

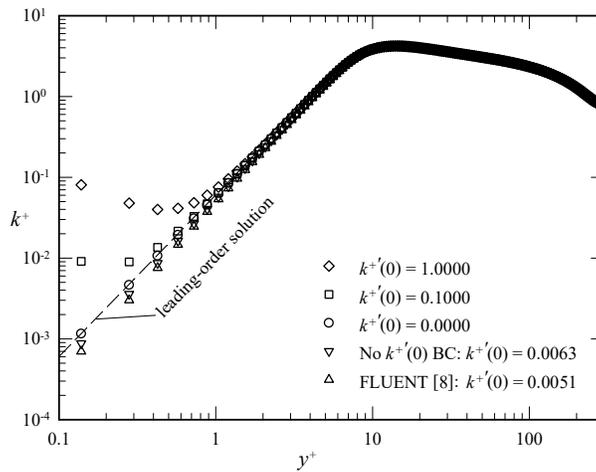

Fig. 10   Effects of wall boundary conditions on turbulent energy predicted from the Wilcox 1998 $k$-$\omega$ model.

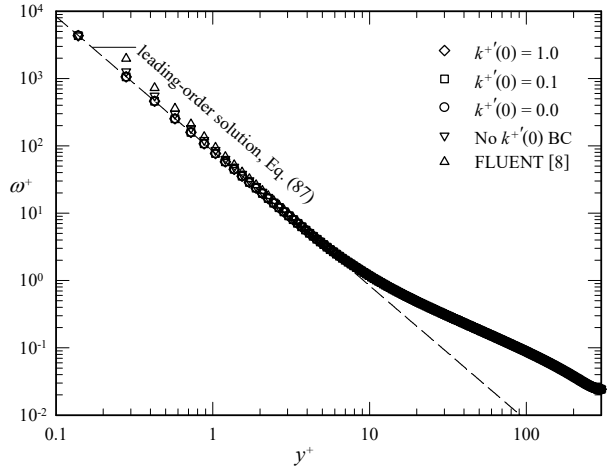

**Fig. 11    Effects of wall boundary conditions on the turbulent dissipation frequency predicted from the Wilcox 1998 $k$-$\omega$ model.**

Like the Lam-Bremhorst $k$-$\varepsilon$ model, the wall damping functions for the Wilcox 1998 $k$-$\omega$ model decay rapidly with increasing $y^+$, so the wall boundary condition $k^{+'}(0) = 0$ has little impact on the predicted velocity profiles. Furthermore, as recommended by Wilcox [15], the use of smooth-wall boundary conditions can be avoided completely with the $k$-$\omega$ model by using the rough-wall boundary conditions first suggested by Saffman [17]. For a smooth wall, Wilcox recommends using the rough-wall boundary conditions with a finite wall-scaled dimensionless roughness height less than 5. As Wilcox points out, the ability to implement rough-wall boundary conditions is a key advantage of the $k$-$\omega$ formulation over the $k$-$\varepsilon$ formulation. For the most recent advancements in the $k$-$\omega$ model, including wall boundary conditions for rough and *hydraulically smooth* surfaces, see Wilcox [13,18].

## 6.    Conclusions

Despite the comments by Durbin [6], many of the $k$-$\varepsilon$ turbulence models in common use today are based on incorrect wall boundary conditions, e.g., Eq. (39) or Eq. (46). The correct smooth-wall boundary conditions presented by Durbin [6] and given here in Eq. (47) are seldom used with $k$-$\varepsilon$ turbulence models. Furthermore, the smooth-wall boundary conditions given in Eq. (47) should be used with $k$-$\omega$ and $k$-$\zeta$ turbulence models as well. The eddy-viscosity models used with conventional $k$-$\omega$ and $k$-$\zeta$ turbulence models are obtained directly from the $k$-$\varepsilon$ eddy-viscosity model by using simple changes of variables. Hence, applying a wall boundary condition to $\omega$ or $\zeta$ is no more correct than applying the equivalent wall boundary condition to $\varepsilon$. The correct smooth-wall boundary conditions for all $k$-$\varepsilon$, $k$-$\omega$, and $k$-$\zeta$ turbulence models are those given by Eq. (47). Physics imposes no smooth-wall boundary condition directly on $\varepsilon$, $\omega$, or $\zeta$. As has always been the case, boundary conditions must be obtained from physics. They cannot be developed solely from the differential equations, and they cannot be selected arbitrarily for convenience of numerical implementation.

Because the zero-gradient boundary condition for $k$ in Eq. (47) has been omitted from many commonly used turbulence models, these models are short one boundary condition and thus are mathematically indeterminate. Because any solution obtained from these models is not unique, numerical solutions obtained from these models may be highly implementation dependent. Furthermore, the authors have observed that solutions obtained from some implementations of these mathematically indeterminate turbulence models can be very sensitive to the grids and initial estimates used to obtain the solutions. Although one numerical implementation may converge to a solution that agrees closely with the unenforced boundary condition, another implementation of the same turbulence model could converge to a different solution. The fact that some particular numerical implementation converges to a solution that agrees closely with the unenforced boundary condition cannot be used to justify the implementation of any turbulence model that is mathematically indeterminate. Turbulence models must always be implemented with a complete set of physically correct boundary conditions. None of these boundary conditions can ever be replaced with a mathematical relation that has been developed solely from the differential equations.

Many low-Reynolds-number turbulence models, such as the Lam-Bremhorst $k$-$\varepsilon$ model and the Wilcox 1998 $k$-$\omega$ model, have wall damping functions that decay rapidly with increasing $y^+$, so the smooth-wall boundary condition $k^{+'}(0) = 0$ has little effect on the predicted velocity profiles. However, all turbulence models should be implemented in such a way that is mathematically determinate.